\documentclass[aps,prb,reprint,superscriptaddress,amsmath,amssymb,floatfix]{revtex4-2}

\usepackage{xcolor}
\usepackage{graphicx}
\usepackage{booktabs}
\usepackage[scr=boondoxo]{mathalpha}
\usepackage[unicode,colorlinks,allcolors=blue]{hyperref}
\usepackage[normalem]{ulem} 

\renewcommand{\v}[1]{\textbf{#1}}
\newcommand{\dd}[1]{\text{d}{#1}\,}
\newcommand{\ddd}[2]{\text{d}{#1}\text{d}{#2}\,}
\newcommand{\braket}[1]{\langle{#1}\rangle}

\begin{document}

\title{Parafermions in polarized $\nu=2/3$ Fractional Quantum (Anomalous) Hall Heterostructures}

\author{Steffen Bollmann}\thanks{corresponding author: s.bollmann@fkf.mpg.de}
\affiliation{Max-Planck Institute for Solid State Research, 70569 Stuttgart, Germany}

\author{Andreas Haller}
\affiliation{Department of Physics and Materials Science, University of Luxembourg, L-1511 Luxembourg, Luxembourg}

\author{Thomas L. Schmidt}
\affiliation{Department of Physics and Materials Science, University of Luxembourg, L-1511 Luxembourg, Luxembourg}

\author{Jukka I. V\"ayrynen}
\affiliation{Department of Physics and Astronomy, Purdue University, West Lafayette, Indiana 47907, USA }

\author{Elio J. K\"onig}
\affiliation{Department of Physics, University of Wisconsin-Madison, Madison, Wisconsin 53706, USA}

\date{\today}

\begin{abstract}
Motivated by recent experimental signatures of fractional Chern insulators in 2D quantum materials, we investigate the realization of $\mathbb Z_3$ parafermion zero modes (PZMs) on edges of spin-polarized $\nu=2/3$ fractional quantum Hall states which are proximitized by superconductors.  Unlike the spin-unpolarized Halperin state, the polarized $\nu=2/3$ edge contains counterpropagating electron and fractional quasihole modes, giving rise to a plethora of competing tunneling and pairing processes. We focus on the vicinity of the fixed point where these modes reorganize into decoupled charge and neutral sectors, with the neutral sector describing an emergent dipole degree of freedom. Using bosonization and renormalization group analysis, we lay out the parameter space in which the interedge dipole interactions gap the neutral sector to stabilize one dominant tunneling and one dominant pairing channel resulting in a double sine-Gordon model in the charge sector which supports $\mathbb{Z}_3$ PZMs at domain walls. We discuss both clean and disordered systems. In clean systems this mechanism requires a certain degree of control of the interaction strength and range. In contrast, disorder can provide a more generic route: 
While short-ranged, edge-uncorrelated disorder  
suppresses the simultaneous pinning required for stable parafermion modes, 
when the disorder on the two edges is sufficiently correlated, as expected for smooth disorder potentials, the parameter regime hosting PZMs increases.

\end{abstract}

\maketitle

\section{Introduction}

The realization of topologically ordered phases and the associated detection of exotic anyonic quantum particles remains a major focus of modern condensed matter physics \cite{NakamuraManfra2020, NakamuraManfra2023, SureshManfra2025, GuerreroSuarezManfra2025, WerkmeisterKim2025,
Kim2024,
Kim2026,
Samuelson2026,Bartolomei2020,
Glidic2023,
Ruelle2023,
Lee2023,
Kundu2023,
Ghosh2025}. Beyond its inherent scientific value, this endeavor is fueled by the dream of an intrinsically robust topological quantum computer. In this framework, quantum information is encoded in anyonic fusion channels whilst quantum braids are the physical processes underlying quantum gates. \cite{NayakDasSarma2008, Kitaev2003} A famous example includes Majorana-based topological quantum computation \cite{Alicea2012, Microscoft2025, Microsoft2026},  corresponding to Ising topological order. In this case, however, the set of quantum gates is too restricted to enable universal gate operations. As a possible remedy, parafermionic generalizations of Majorana zero modes to fractionalized systems with richer $\mathbb{Z}_n$ degrees of freedom have been put forward \cite{FradkinKadanoff1980, ZamolodchikovFateev1985, Fendley2014, MongFendley2014}. In suitable coupled architectures, they can provide a route toward Fibonacci anyons -- a platform for universal topological quantum computation~\cite{Mong2014, Vaezi2014a,AliceaFendley2016}. 

Parafermion zero modes (PZMs) have been proposed to emerge at interfaces between differently gapped regions of fractional quantum Hall (FQH) edges ~\cite{LindnerStern2012, ClarkeAliceaShtengel2013,Cheng2012}. The topic has received substantial theoretical attention and has been extensively studied \cite{Mong2014, Vaezi2013, Vaezi2014a, ClarkeAliceaShtengel2014, CobaneraOrtiz2014, SantosHughes2017,LiangLyanda-Geller2019,Schmidt2020, LiCheng2015, EbisuOreg2017,CalzonaSchmidt2018, SnizhkolGefen2018a, SnizhkoGefen2018b, GroenendijkCalzonaSchmidt2019,NielsenBurrello2022,CaoFradkin2024, YoshitomeGomes2025, ZhouLyanda-Geller2026, BollmannKoenig2026, CaoFradkin2026}.  
A major experimental challenge is the simultaneous realization of a robust fractional topological phase and superconductivity \cite{GuelKim2022, SchillerOreg2023, HuWangLian2024,KurilovichGlazman2023, ZhaoFinkelstein2023,UdayAndo2024, TamLian2026}. Conventional FQH states require strong magnetic fields, which complicate their integration with superconductors (SCs). Fractional Chern insulators (FCIs) provide an attractive alternative: they realize fractionalized topological order in partially filled Chern bands and can occur without an externally applied magnetic field~\cite{RegnaultBernevig2011, NeupertMudry2011, LiuBergholz2024}. Their recent observation in 
2D quantum materials
\cite{Cai2023,ZengShan2023,ParkXu2023,XuLi2023,JiShen2024,RedekopYoung2024,LuJu2024,ChenWang2024,LuJu2025,DingXu2025}, therefore, opens new possibilities for realizing parafermions in experimentally accessible platforms.

\begin{figure}
    \centering
    \includegraphics[width=\linewidth]{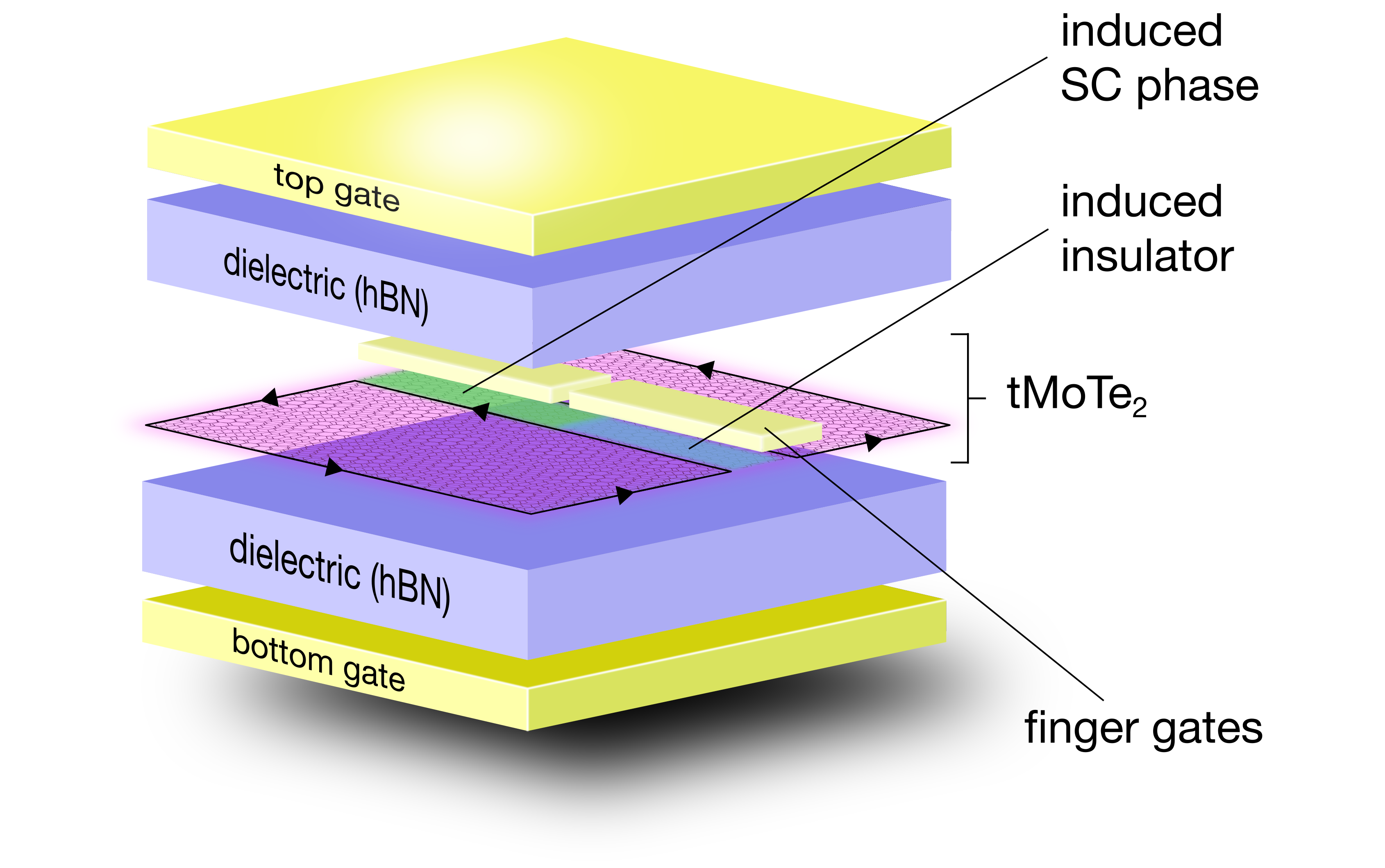}
    \caption{Schematic of hypothetical experimental setup. A sample of  tMoTe$_2$ is tuned into a $\nu=2/3$ FQH state (purple glow) via tuning the filling with a top and bottom gate. Within the FQH state, a thin finger gate induces superconducting (green) and insulating regions (purple) within the sample.}
    \label{fig:experiment}
\end{figure}

Twisted bilayer $\mathrm{MoTe}_2$ (tMoTe$_2$) is particularly promising in this respect. Robust fractional states have been observed at filling $\nu=2/3$, while superconductivity has recently been reported to appear within the same moir\'e Chern band, slightly doping away from the fractional regime~\cite{XuLi2023}. This reignited the theoretical study of anyon-driven superconductivity  \cite{Laughlin1988, ChenLaughlin1989, FetterHannaLaughlin1989, FanLi2025,ShiSenthil2025,DivicVishwanath2025,HanWangVishwanath2025,NosovHanKhalaf2026,PichlerVishwanath2026,Goldman2026,MehtaGoldman2026,SeoYoung2026} and at the same time raises the possibility of realizing an FCI--SC heterostructure within a single material. A schematic setup inspired by similar previous studies \cite{LeeKim2017, GuelKim2022, JiaWu2024, HanJu2025, XuLi2025} is shown in Fig.~\ref{fig:experiment}. Top and bottom gates tune the sample into the spin-polarized $\nu=2/3$ FCI, while a narrow finger gate locally changes the filling and induces superconducting or insulating regions. Fractional superconducting and insulating regions can thus, in principle, be defined electrostatically without an additional material interface.

With the prospect of such a single-material realization, we here theoretically delineate the parameter space in which $\mathbb Z_3$ parafermion modes may be realized based on 1D heterostructures of SCs and the edge states of spin-polarized $\nu = 2/3$ phases. Our study includes an in-depth analysis of interaction and disorder effects. We find that for clean samples, interaction effects need to be 
externally controlled
to stabilize quantum phases with parafermion edge states. In contrast, for disordered interfaces, the
tuning can be avoided. 
If the disorder potential is long ranged (e.g. Coulomb impurities), the parafermionic phase is stable over a large portion of the phase diagram. However, if the disorder potential is very short ranged and at the same time the tunneling interface is smooth, the phases including parafermions are suppressed.

This manuscript is structured as follows. As a technical motivation, we start in Sec.~\ref{sec:recap} by giving a brief review of how PZMs can appear in FQH--SC heterostructures as a low-energy theory of a double sine-Gordon model. Thereby, we identify the unresolved question of $\mathbb Z_3$ PZMs in a $\nu=2/3$ spin-polarized FQH state.
In Sec.~\ref{sec:microscopicModel}, we introduce the model and define all relevant objects needed to derive the effective low-energy theory. In Sec.~\ref{sec:cleanCase}, we first demonstrate our approach in a fine-tuned case in which the edge reorganizes into two counterpropagating decoupled modes, one neutral and one carrying charge $2e/3$. Subsequently, in Sec.~\ref{sec:dirtyCase}, we discuss the more realistic case of correlated and uncorrelated edge disorder for multiple disorder scenarios and demonstrate a $\mathbb Z_3$ PZM phase without the necessity of fine-tuning. Finally, in Sec.~\ref{sec:Conclusion}, we summarize and discuss our results.
\section{Parafermions in FQH--SC heterostructures}
\label{sec:recap}

We briefly review how fractional quantum Hall -- superconductor (FQH-SC) heterostructures can act as a platform to realize parafermions. In the following, we follow the strategy laid out in Ref.~\cite{ClarkeAliceaShtengel2013}. Throughout the manuscript, we use units $\hbar = k_B = 1$ in which both the reduced Planck constant and Boltzmann's constant are set to unity.

\begin{figure}
    \centering
    \includegraphics[width=\linewidth]{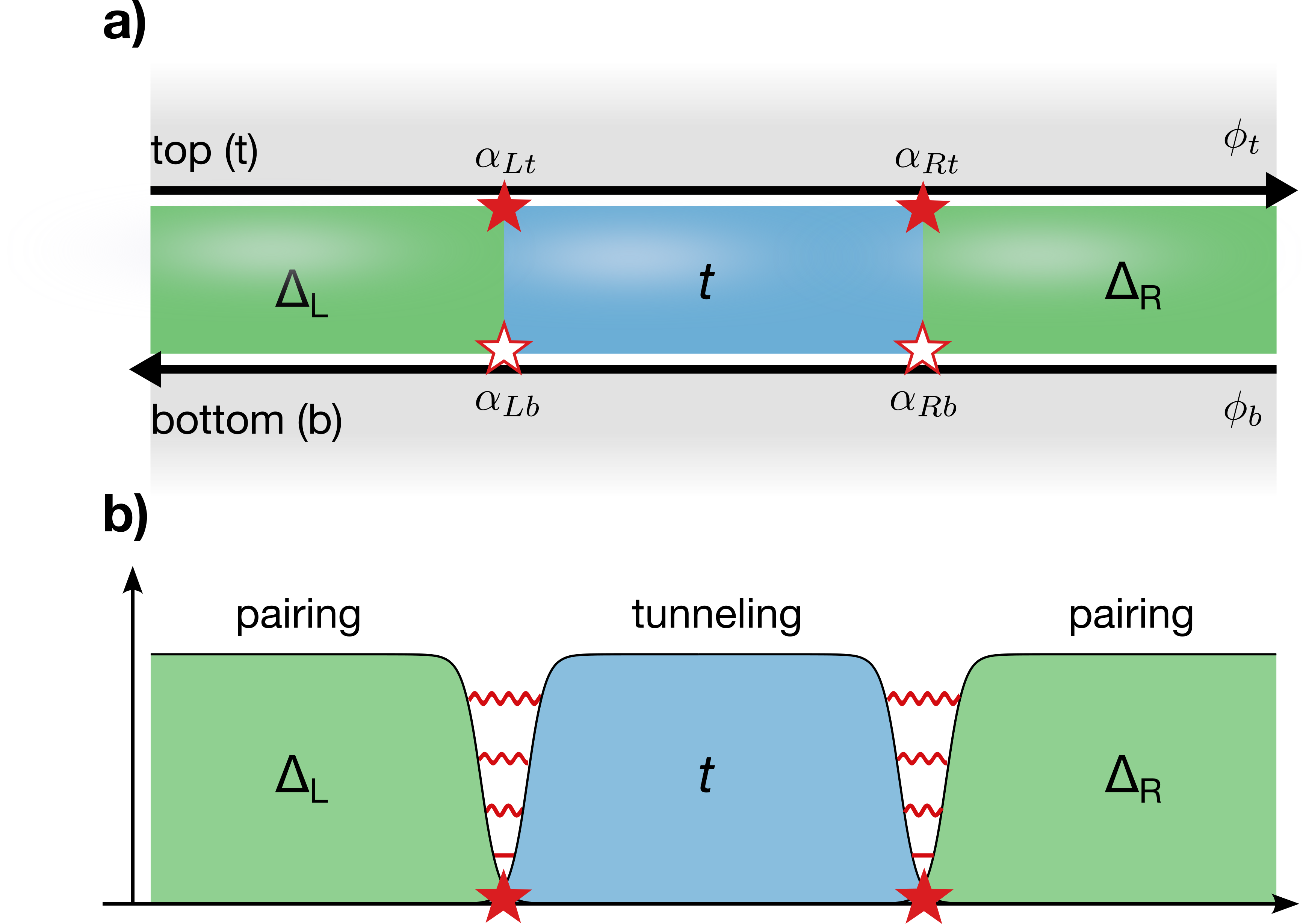}
    \caption{\textbf{a)} Schematic of a FQH-SC heterostructure. A trench in an FQH liquid is subsequently filled with a SC ($\Delta_L$), an insulator ($t$), and a SC ($\Delta_R$). At the interfaces, parafermions (red stars) emerge. \textbf{b)} Potential landscape due to tunneling and pairing. The interfaces create potential wells in which bound states can be trapped. The zero-energy states (red stars) have parafermionic statistics.}
    \label{fig:RecapFigure}
\end{figure}

\subsection{Fractional quantum Hall edge states}

Let two fractional quantum Hall edges at filling factor $\nu=1/k$ be separated by a thin trench. The trench is filled with a SC 
(thinner than its coherence length) and regions of insulating material that allow for electron pairing, i.e., crossed Andreev reflection (CAR) and tunneling, respectively. These materials form interfaces at $x_R$ and $x_L$ as shown Fig.~\ref{fig:RecapFigure}.

On the FQH edges, the two edge modes (``top'' and ``bottom'') can be described by two counterpropagating bosonic modes with the usual commutation relations
\begin{subequations}
    \begin{align}
        [\phi_t(x), \phi_t(x')] & = \frac{i\pi}{k}\text{sign}(x-x'), \\
        [\phi_b(x), \phi_b(x')] & = -\frac{i\pi}{k}\text{sign}(x-x'), \\
        [\phi_t(x), \phi_b(x')] & = -\frac{i\pi}{k},
    \end{align}
\end{subequations}
whose dynamics in the absence of CAR or tunneling are determined by the quadratic Hamiltonian 
\begin{equation}
    H_0 = \sum_{\tau=t,b} v_\tau \int \dd x  (\partial_x\phi_\tau)^2.
\end{equation}
Note that the non-vanishing commutator between top and bottom edge modes presumes that these modes are connected at $x \to +\infty$ and avoids the introduction of Klein factors~\cite{GuyonSafi2002}.

A trench of SC or insulator separates the top and bottom edge modes, see Fig.~\ref{fig:RecapFigure}a).
Only electrons can cross the trench, which can be represented in terms of vertex operators of top and bottom modes,
\begin{equation}
    \psi_\tau \sim e^{ik\phi_\tau}, \quad \tau = t,b.
\end{equation}
The corresponding density operators are $\rho_{t/b} = \pm\frac{1}{2\pi} \partial_x\phi_{t/b}$. It is useful to introduce the non-chiral fields
\begin{equation}
    \Phi = \frac{1}{2}(\phi_t +\phi_b)\quad\text{and}\quad\Theta=\frac{1}{2}(\phi_t - \phi_b),
\end{equation}
where $\Phi$ is the phase canonically conjugate to the total density $\rho=\frac{1}{\pi}\partial_x\Theta$. Their mutual commutation relation reads
\begin{equation}
    [\Phi(x), \Theta(x')]=i\frac{\pi}{k}\theta(x-x'),
    \label{eq:commutationRelationPhiTheta}
\end{equation}
with $\theta(x-x')$ being the Heaviside step function.

Bosonizing the tunneling and pairing terms, 
\begin{equation}
\begin{split}
     H & = - \int  \dd x t(x)  \psi_t^\dagger\psi_b +  \int\dd x \Delta(x) \psi_t \psi_b  + \text{H.c.} \\
    & \sim \int  \dd x t(x) \cos(2k\Theta) - \int\dd x \Delta(x)\cos(2k\Phi),
    \label{eq:reviewBosonizedTunnelingPairing}
\end{split}
\end{equation}
eventually yields a Hamiltonian that 
resembles a double sine-Gordon model \cite{Lecheminat2002}. 
We will assume regions with topologically distinct gaps corresponding to regions with predominant tunneling and predominant pairing forming an interface. For theoretical simplicity,
the support of the functions $t(x)$ and $\Delta(x)$ is  assumed to have zero overlap as displayed in Fig.~\ref{fig:RecapFigure}b).

\subsection{Semiclassical minima and parafermionic bound states}
\label{sec:MinimaAndDerivationOfParafermions}

According to Hamiltonian \eqref{eq:reviewBosonizedTunnelingPairing}, the fields will be pinned to the values $\Theta = \frac{\pi}{k}\hat m-\frac{\pi}{2 k}$ and $\Phi = \frac{\pi}{k}\hat n_{L,R}$ in the tunneling and left/right pairing region, respectively. Here, $\hat m$ and $\hat n_{L,R}$ are integer-valued operators whose eigenvalues depend on the particular minimum in which the fields $\Theta$ and $\Phi$ are pinned. They inherit the commutation relation 
\begin{subequations}
\begin{align}
    [\hat n_R, \hat m ] &=i\frac{k}{\pi}, \\
    [\hat n_L, \hat m] & = 0
\end{align}
\label{eq:nmCommutation}
\end{subequations}
directly from Eq.~\eqref{eq:commutationRelationPhiTheta}.

As illustrated in Fig.~\ref{fig:RecapFigure}a), the interfaces form potential wells capable of trapping bound states. These states must satisfy the boundary conditions imposed by the pinned $\Theta$ and $\Phi$ fields in the tunneling- and pairing-dominated phases, respectively. At the interface, localized zero-energy modes can be constructed. For temperatures below the energy gap to the first excited states, the projected zero-energy states, indicated by the red stars in Fig.~\ref{fig:RecapFigure}, can then, up to phase factors, be expressed as
\begin{align}
\begin{split}
    \alpha_{tR} \sim e^{i\frac{\pi}{k}(\hat n_R + \hat m)}, \quad
    \alpha_{tL} \sim e^{i\frac{\pi}{k}(\hat n_L +\hat m)},  \\
    \alpha_{bR} \sim e^{i\frac{\pi}{k}(\hat n_R - \hat m)}, \quad 
    \alpha_{bL} \sim e^{i\frac{\pi}{k}(\hat n_L -\hat m)}.
\end{split}
\end{align}
Note that the two pairs $\{\alpha_{tR},\alpha_{tL}\}$ and $\{\alpha_{bR},\alpha_{bL}\}$ of parafermions are not independent, as they only differ by a phase shift. To indicate that the Hilbert space is only spanned by one of the two operator sets, the stars representing $\{\alpha_{bR},\alpha_{bL}\}$ in Fig.~\ref{fig:RecapFigure}a) are empty.

Together with the commutation relation \eqref{eq:nmCommutation} one can show that the zero modes $\alpha_{\tau i}$ fulfill the $\mathbb Z_{2k}$ parafermion algebra
\begin{equation}
    (\alpha_{\tau i})^{2k}=1,\quad \alpha_{\tau L}\alpha_{\tau R} =\alpha_{\tau R}\alpha_{\tau L}e^{-i\frac{\pi}{k}}. 
\end{equation}
In this sense, we define a realization of parafermions in the given FQH–SC heterostructure as the existence of a low-energy theory that pins the dual fields $\Theta$ and $\Phi$ in distinct spatial regions separated by an interface.

We conclude this section with a comment on the interface between the one-dimensional SC and the FQ(A)H insulator (i.e., replacing the region labelled $t$ in Fig.~\ref{fig:RecapFigure}a) by the FQ(A)H insulator such that the edge states form a U-turn hugging the green $\Delta$-regions. It turns out that this situation can be obtained from our setup in the strong tunneling limit. Note that, effectively, strong tunneling can also be reached for weak bare tunneling amplitudes, if the residual interactions turn the coupling constant $t$ to be relevant. From the perspective of the quantum Hall state~\cite{Haldane1994}, in the infrared, the junction has become transparent to transporting fractional charges~\cite{KaneFisher1992}. We refer to this phenomenon as ``healing'' of the trench.
As a corollary of relevance for the physics discussed in this paper, the PZMs are theoretically expected also at the end of a superconducting finger sticking into the FQ(A)H liquid -- as experimentally realized in Ref.~\cite{GuelKim2022} -- provided that at least $\Delta$ is a relevant perturbation.

\subsection{Parafermions beyond Laughlin states}

\begin{figure}
    \centering
    \includegraphics[width=\linewidth]{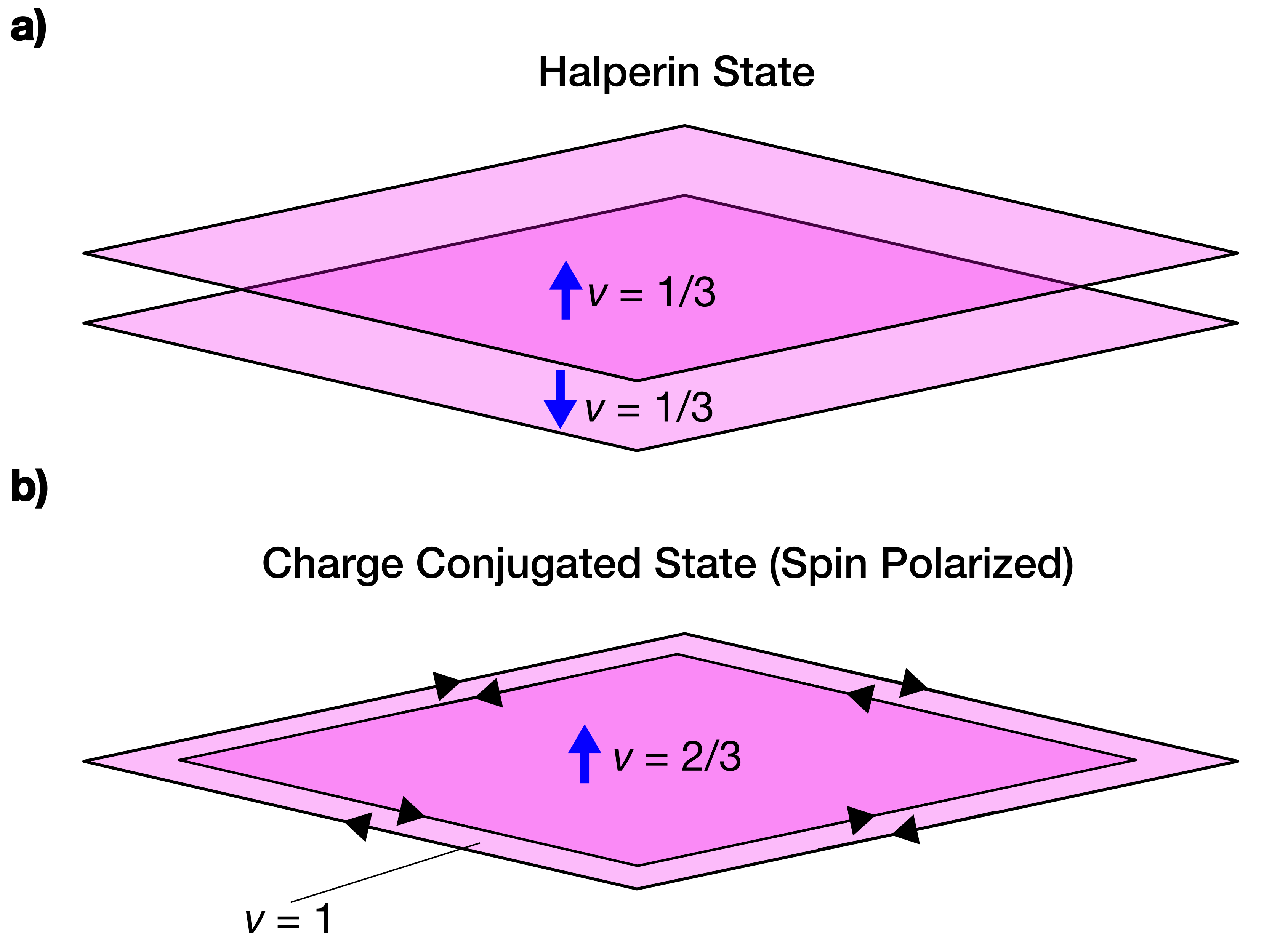}
    \caption{a) Two quantum Hall liquids at filling $\nu=1/3$ with opposite spin polarization are coupled to form an SU(2)-symmetric $\nu=2/3$ Halperin (112) state. 
    b) Spin polarized charge conjugate $\nu=2/3$ state. The edge consists of two counterpropagating edge modes.}
    \label{fig:FQHStates}
\end{figure}

While the previous derivation, using Laughlin states, seems to only realize $\mathbb Z_{2k}$ parafermions, it has also been shown that a similar construction can be carried out in the presence of more complex states. For instance, the FQH states with filling fractions $\nu=2/3$ can realize a double sine-Gordon model with three-fold periodicity (i.e., replacing $``2k"$ by $``3"$ in the trigonometric functions of Eq.~\eqref{eq:reviewBosonizedTunnelingPairing}) and thus $\mathbb Z_3$ parafermions. There are multiple different realizations of the FQH states with filling fraction $\nu=2/3$ that realize the same topological order but differ significantly in their physical implications at the edge theory. The (spin-unpolarized) $\nu=2/3$ Halperin 112 state \cite{Halperin1983} describes two coupled 
quantum Hall fluids with opposite spin polarizations (see Fig.~\ref{fig:FQHStates}a), which are often considered SU(2) symmetric \cite{Mong2014, Vaezi2014a}.  
Another $\nu=2/3$ state, relevant to spin-polarized Landau levels and non-degenerate Chern bands and believed to be realized in twisted MoTe$_2$ or rhombohedral graphene, is the particle–hole conjugate of the $\nu=1/3$ Laughlin state. As illustrated in Fig.~\ref{fig:FQHStates}, this state can be understood as a $\nu=1/3$ Laughlin liquid of holes in an otherwise filled Landau level or Chern band \cite{Laughlin1983, Girvin1984}. Its edge hosts two counterpropagating modes, an integer mode supporting electron excitations of charge $-e$ and a fractional mode supporting quasiparticle excitations of charge $e/3$, which are separated by the distance $r$, where $e>0$ is the elementary charge. The latter state can occur spin-polarized or spin-unpolarized \cite{XieGuoZhang1989, EisenesteinWest1990}. However, since the FCI state realized in twisted MoTe$_2$ as well as rhombohedral multilayer graphene is most likely spin-polarized \cite{WangXiao2024, DongSenthil2024}, we focus in this work only on the polarized case. 

To understand this in more detail, it is worthwhile to revisit the construction of the electron from FQH edges via vertex operators. Let a general FQH edge host $\boldsymbol{\phi} =(\phi_{k_1},\phi_{k_2},\dots,\phi_{k_N})$ edge bosons. Here, the $k_j$ are elements of some index set. 
If all fields have a unit compactification radius, a generic quasiparticle is defined by the integer-valued vector $\v l$ via $V_{\v l }\sim e^{i\v l\cdot\boldsymbol{\phi}}$. For $V_{\v l}$ being an electron, it must create a quasiparticle of charge $-e$, and have a statistical angle $\theta=n\pi$, with $n$ being odd, such that the electronic vertex operators anti-commute. These constraints still allow infinitely many $\v l_j$ that fulfill these criteria. Thus, the physical electron can be constructed from a superposition of all these possible electron operators, that is \cite{WenBook2007}
\begin{equation}
    \psi = \sum_{j}c_j e^{\v l_{j}\cdot \boldsymbol \phi},
    \label{eq:electronOperatorGeneral}
\end{equation}
where the coefficients $c_j$ are determined by microscopic details. While all $V_{\v l_j}$ have the same physical charge and fermionic exchange statistics, their scaling dimensions are in general different. Thus, if one is only interested in the low-energy theory, one may only keep the most relevant operators in the sum \eqref{eq:electronOperatorGeneral}.

In the Halperin state, since the edge theory is symmetric under flipping the spin, one can construct two equally relevant fermionic operators. One corresponds to a spin-up and the other to a spin-down electron. Considering spin-conserving tunneling and singlet pairing, bosonization results in a double sine-Gordon model of the shape in Eq.~\eqref{eq:reviewBosonizedTunnelingPairing} \cite{Mong2014, Vaezi2014a}. 

The situation becomes more complicated once one considers the spin-polarized state. This is because there is no underlying spin-SU(2) symmetry. For instance, perturbing around the decoupled electron $\phi_{\tau1}$ and $e/3$-quasi-hole $\phi_{\tau1/3}$ edge theory, the two most relevant electronic vertex operators are 
\begin{equation}
    \psi_{\tau1} \sim e^{i\phi_{\tau1}}\quad\text{and}\quad \psi_{\tau1/3} \sim e^{i3\phi_{\tau1/3}},
\end{equation}
where $\Delta_1 =\frac{1}{2}$ and $\Delta_{1/3}=\frac{3}{2}$ are the scaling dimensions of the respective operators in the absence of residual interactions. Thus, for small interaction around the fixed point of decoupled electron and quasi-hole edge modes, the largest low-energy contribution comes from the electron edge modes, which would yield Majorana fermions at the interface, but no parafermions. On the other hand, if the interactions are strong such that $\Delta_{1/3}<\Delta_1$, 
only the $1/3$ mode would contribute, and the interface traps $\mathbb Z_6$ parafermions. This illustrates 
that only a non-trivial combination of several electron operators can actually give rise to $\mathbb Z_3$ parafermions in analogy to the Halperin state. 
However, it is not a priori clear which tunneling and pairing channels of these multiple 
electron operators will contribute most and must be handled simultaneously. 

In the remainder of the paper, we resolve these conceptual barriers by considering the vicinity of a fixed point at which the electron and quasihole modes reorganize into a charge mode with charge $-2e/3$ and a neutral mode, and scrutinize the parameter regime hosting PZMs. At this \textit{Kane-Fisher-Polchinski} (KFP)~\cite{KaneFisherPolchinski1994} fixed point, the two modes are decoupled and do not interact with each other. Here, this fixed point can be reached by two means: For clean samples, interactions need to be fine-tuned along a line of fixed points (for the stability of PZMs, exact fine-tuning is not necessary, see Sec.~\ref{sec:cleanCase} below). Alternatively, the decoupled fixed point is stabilized (i.e., it is an attractive fixed point) in the presence of
disorder-induced quasiparticle backscattering. The attractive 
KFP fixed point in the presence of disorder was originally proposed to explain the experimentally observed robust two-terminal conductance of $G=\frac{2}{3}\frac{e^2}{h}$, in contrast to an expected non-universal conductance $G$ for a clean system of
interacting electron and quasihole modes. 

Starting from the vicinity of this fixed point, we derive a Hamiltonian of the shape of Eq.~\eqref{eq:reviewBosonizedTunnelingPairing} as the appropriate low-energy theory and demonstrate that, through the construction outlined above, $\mathbb Z_3$ parafermions can be trapped at the interface between tunneling and superconducting regions. We first consider the clean but fine-tuned case to demonstrate the mechanism and then discuss the more generic case of disordered edges and how the disorder influences the arguments laid out in Sec.~\ref{sec:MinimaAndDerivationOfParafermions}. 
Our results reveal a sizable region of parameter space in which PZMs are stabilized.

\section{Microscopic Model}
\label{sec:microscopicModel}

\begin{figure}
    \centering
    \includegraphics[width=\linewidth]{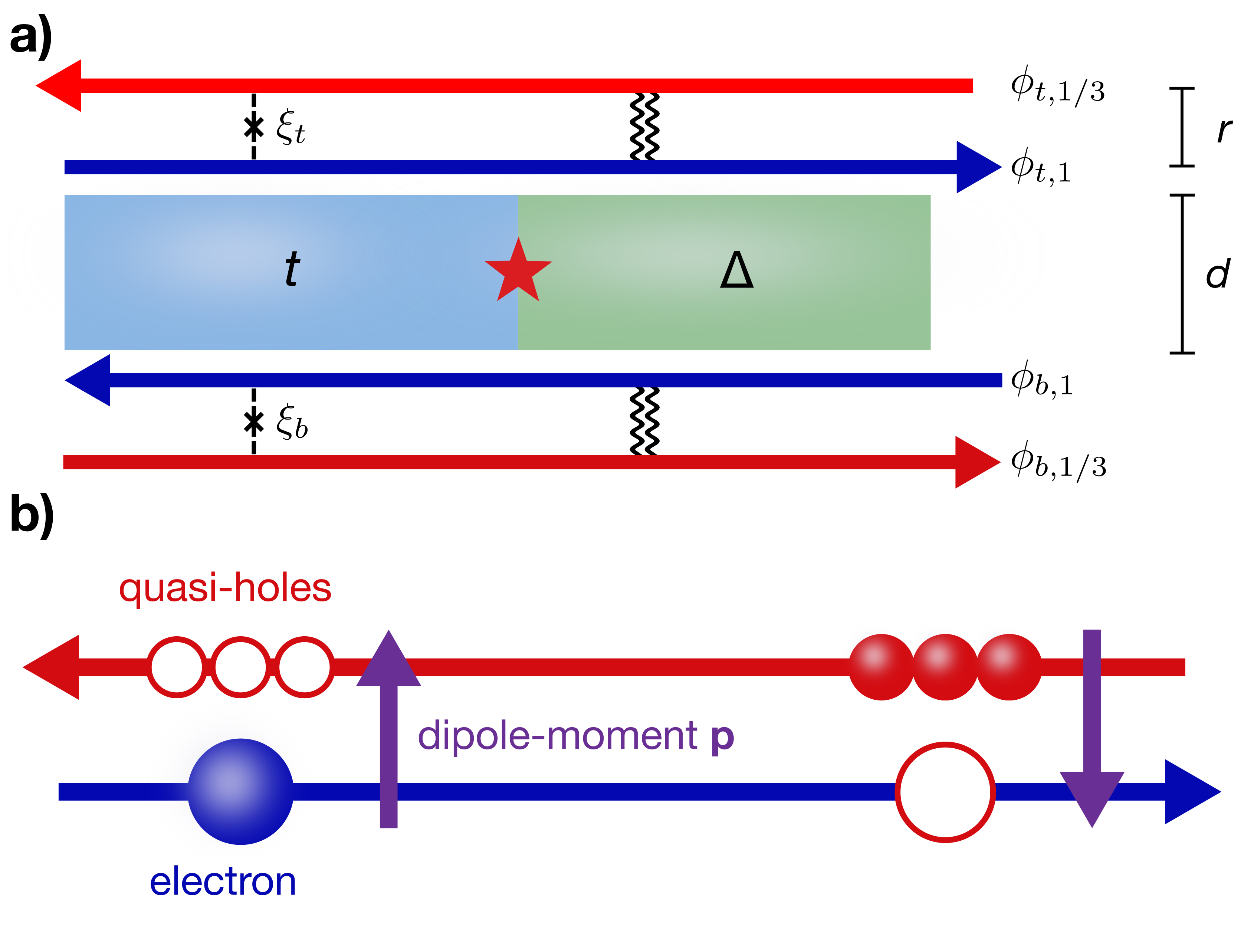}
    \caption{\textbf{a)} Schematic of the interface of a tunneling (t) and pairing ($\Delta$) region coupled to two $\nu=2/3$ spin-polarized FQH states. The top and bottom edges consist of two counterpropagating edge modes, one carrying charge $-e$ (blue) and one carrying charge $e/3$ (red). At each edge, the modes are separated by a distance $r$. The two edges themselves are separated by the distance $d$. \textbf{d)} Internal structure of the neutral dipole moments (purple arrow). An ``up" dipole consists of three quasiholes and one electron. A ``down" dipole is an electron hole together with three charge $e/3$ quasiparticles.}
    \label{fig:UVModel}
\end{figure}

In this section we recapitulate the general edge structure of the $\nu=2/3$ spin polarized FQH state and derive a microscopic model that describes the heterostructure as shown in Fig.~\ref{fig:UVModel}. Thereby, we discuss all the important objects such as electron operators represented as vertex operators of the bosonic edge modes and an emergent $\mathfrak{su}(2)_1$ algebra that describes physical charge dipole moments.

\subsection{Ultraviolet Edge Theory}

The canonical starting point for the edge theory of a spin-polarized $\nu=\frac{2}{3}$ fractional quantum Hall state consists of two counterpropagating bosonic edge modes: an electronic mode $\phi_1$ carrying charge $-e$ and a fractional quasihole mode $\phi_{1/3}$ carrying charge $e/3$ \cite{FradkinBook2013, WenBook2007} with the commutation relations 
($p,q \in \{1,1/3\}$)
\begin{subequations}
\begin{align}
    [\phi_{t,p}(x),\phi_{t,q}(x')]
    &=
    i\pi(K^{-1})_{pq}\,\text{sign}(x-x'),
    \label{eq:CommutatorTop}\\
    [\phi_{b,p}(x),\phi_{b,q}(x')]
    &=
    -i\pi(K^{-1})_{pq}\,\text{sign}(x-x'),
    \label{eq:CommutatorBottom}\\
    [\phi_{t,p}(x),\phi_{b,q}(x')]
    &=
    -i\pi(K^{-1})_{pq}.
    \label{eq:CommutatorMixed}
\end{align}
dictated by the ${K}$ matrix
\begin{align}
    K = \begin{pmatrix}
        1 & 0\\
        0 & -3
    \end{pmatrix}.
    \label{eq:bareKmatrix}
\end{align}
\end{subequations}

We consider the setup shown in Fig.~\ref{fig:UVModel}b, where the electronic and quasi-hole modes are separated by a distance $r$. The top $(t)$ and bottom $(b)$ edges, separated by a distance $d$, are coupled through a tunneling-dominated region for $x<0$ and a CAR-dominated region for $x>0$, forming an interface at $x=0$.

The microscopic Hamiltonian
reads
\begin{subequations}
\begin{equation}
    H = H_\text{kin}+ 
    H_\text{int} + H_t + H_\Delta + H_\text{dis}, \label{eq:fullHamiltonian}
\end{equation}
where 
\begin{align}
    H_\text{kin} & = \sum_{\tau,p}v^{(0)}_p \int \dd x (\partial_x\phi_{\tau p})^2, \\
    H_\text{int} & = \sum_{\tau\tau'= t,b}\int \text{d}x\text{d}x'\text{d}y\text{d}{y'} \Big [U_{\tau\tau'}(x,x',y,y') \notag \\
    &\times \rho_{\tau}(x,y)\rho_{\tau'}(x', y') \Big ], 
    \label{eq:HintInUV}\\
    H_t & = -\int_{-\infty}^0 \dd x t  (\psi^\dagger_t\psi_b + \text{H.c.}), \\
    H_\Delta &  = \int_{0}^\infty\dd x \Delta  (\psi_t \psi_b + \text{H.c.}), \\
    H_\text{dis} & = \frac{1}{a}\sum_{\tau=t,b}\int\dd x \left(\xi_\tau(x) e^{i(\phi_{\tau1}+3\phi_{\tau1/3})}+\text{H.c.}\right).
    \label{eq:DisorderedHamiltonian}
\end{align}
\label{eq:UVFullHamiltonian}
\end{subequations}
Here, $a$ is a short-distance cut-off.
In the subsequent sections, we elaborate in detail on each term of this Hamiltonian.

The first term contains the kinetic energy of the bosonic modes with bare velocity $v^{(0)}_p$ predominantly determined by the confining edge potential.

All inter- and intra-edge interactions are encoded in the interaction kernel $U(x,x',y,y')$, which includes electrostatics and all other possible microscopic interaction channels of the charge densities $\rho_\tau(x,y)$. In particular, these density interactions renormalize the mode velocities. In the following, we write $v_{p} = v^{(0)}_p +\delta v_{p,\text{int}}$, where $\delta v_{p,\text{int}}$ contains all {intra-edge interaction} corrections due to Eq.~\eqref{eq:HintInUV}.  

The terms $H_t$ and $H_\Delta$ correspond, respectively, to tunneling and pairing between the two edges, and are written in terms of the physical electron operators.
The physical electron operators $\psi_\tau=\sum_{\{j\}}c_{\tau j}\psi_{\tau j}$  consist of a sum of infinitely many electronic vertex operators \cite{WenBook2007}. The sum may be truncated to only keep the (most) RG-relevant operators, see Sec.~\ref{sec:Elop} for details. 

In addition, we include quenched disorder-induced backscattering, $H_{\rm dis}$, which couples counterpropagating edge modes on each side of the trench through charge-conserving tunneling. Disorder also generates forward scattering, which acts as a spatially varying chemical potential. This contribution can be eliminated from the quadratic Hamiltonian by a position-dependent shift of the bosonic fields.  Random interedge backscattering and/or crossed Andreev reflection will also be considered and will result in an $x$-dependence of $t, \Delta$. The scattering amplitude of intra-edge backscattering is described by the zero-mean Gaussian random variable $\xi_\tau(x)$ with local correlations. In the following, we will discuss three cases: (i) the clean case, where $\xi_\tau=0$, (ii) the case of correlated disorder (e.g., Coulomb impurities) where $\xi_t(x) = \xi_b^*(x)$  and
\begin{align}
\label{eq:DisorderPotential}
    \overline{\xi_\tau^*(x)\xi_{\tau}(x')}
    &=
    W\delta(x-x'),
\end{align} 
as well as (iii)  the case of uncorrelated disorder (e.g., short-range impurities),
where $\xi_t(x), \xi_b^*(x)$ are unrelated {but Eq.~\eqref{eq:DisorderPotential} still holds}. Here, the overline indicates a disorder average, and $W$ is the disorder strength. 

\begin{figure*}
    \centering
    \includegraphics[width=\linewidth]{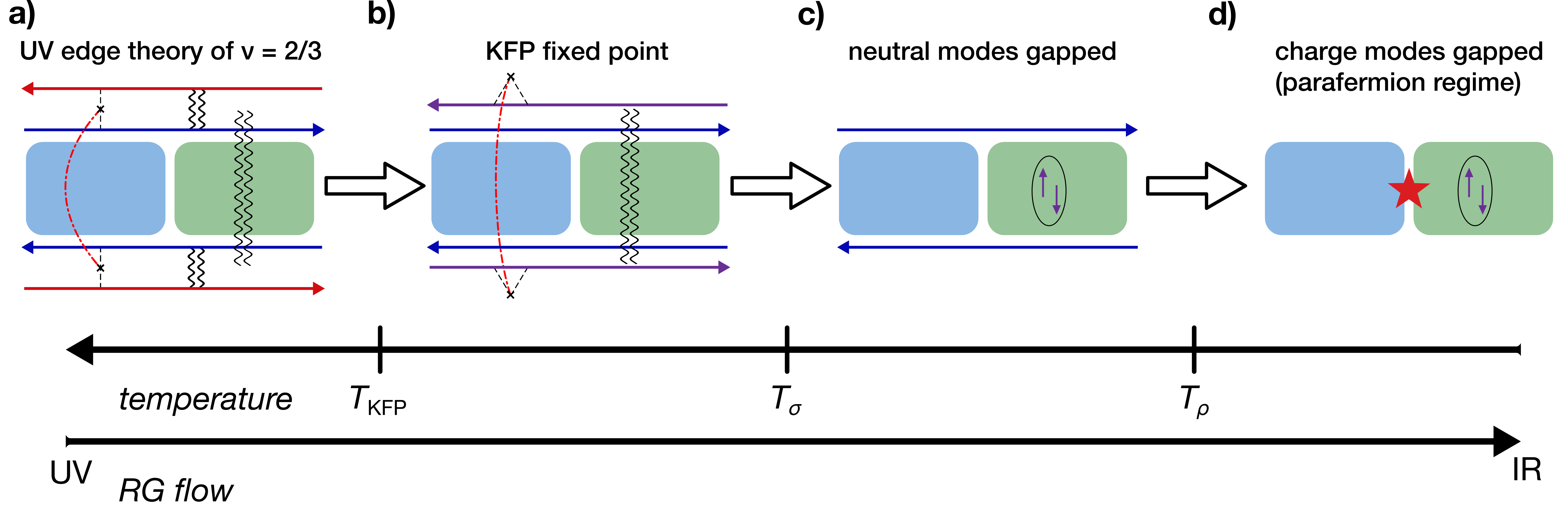}
    \caption{Schematic of the multi-step renormalization group flow and the hierarchy of energy scales. In the presence of disorder-induced backscattering, at $T_\text{KFP}$, the edge reorganizes into a decoupled charge-$2e/3$ and neutral modes. In a second RG step, inter-edge interactions gap out the neutral channel at $T_\sigma$. For edge-correlated disorder, tunneling and pairing channels persist the disorder average and gap out the charged modes, and parafermion zero-energy states emerge. }
    \label{fig:Mechanisim}
\end{figure*}

\subsection{The KFP fixed point in the absence of tunneling and CAR}

Near the KFP fixed point  the edge modes decouple into a neutral $\phi_{\tau\sigma}$ and a charge-$2/3$ $\phi_{\tau\rho}$ modes (either due to fine-tuned interaction $U_{\tau\tau'}$ or driven by disorder)~\cite{KaneFisherPolchinski1994}. Thus, at the temperature scale $T_{\text{KFP}}$, all intra-edge interaction between the $\rho$ and $\sigma$ channels are effectively zero and disorder purely couples to the neutral channel, see Eq.~(\ref{eq:KFPbasis}). This first RG step is depicted in Fig.~\ref{fig:Mechanisim} where the system flows from the UV theory a) to the KFP fixed point b). For fine-tuning in the clean case, $T_\text{KFP}=D_\text{UV}$, where $D_\text{UV}$ is the UV energy scale: 
the edge modes are decoupled throughout all energy scales. 
Far from the fine-tuned point, disorder-induced backscattering is needed to decouple the edges into a charged and neutral mode at a temperature scale $T_\text{KFP}=D_\text{UV}\left(\frac{W}{v_\sigma D_\text{UV}}\right)^{1/(3-2\Delta_W)}$ at which the disorder strength reaches strong coupling, with $v_\sigma$ the velocity of neutral modes (see below). Furthermore, $\Delta_W$ is the scaling dimension of the disorder term Eq.~\eqref{eq:DisorderedHamiltonian}. 

\subsubsection{Generalities}

At the KFP fixed point, the decoupled edge modes in terms of the UV fields are
\begin{subequations}
\label{eq:KFPbasis}
\begin{align}
    \phi_{\tau,\rho}
    &=
    \sqrt{\frac{3}{2}}
    \left(
        \phi_{\tau,1}
        +
        \phi_{\tau,1/3}
    \right),
    \\
    \phi_{\tau,\sigma}
    &=
    \sqrt{\frac{1}{2}}
    \left(
        \phi_{\tau,1}
        +
        3\phi_{\tau,1/3}
    \right).
\end{align}
\end{subequations}
with the corresponding $K$ matrix
\begin{equation}
    K_{\text{KFP}}
    =
    \begin{pmatrix}
        1 & 0\\
        0 & -1
    \end{pmatrix}.
    \label{eq:KFPmatix}
\end{equation}
The commutation relations are identical to Eqs.~\eqref{eq:CommutatorTop}--\eqref{eq:CommutatorMixed}, with $K$ replaced by $K_{\text{KFP}}$ and $p,q\in\{\rho,\sigma\}$.  

These new fields may be interpreted as bound states of the underlying microscopic quasiparticles, namely electrons and quasiholes. In this sense, the charged $2e/3$ mode corresponds to an electron combined with one quasihole, while the neutral mode corresponds to one electron combined with three quasiholes. Thus, although the neutral modes are overall charge neutral, they retain an internal structure due to the microscopic spatial separation of the $\phi_{\tau1}$ and $\phi_{\tau1/3}$ fields. This gives rise to a dipole moment~\cite{BidMahalu2010} $\mathbf p=e\mathbf r,$
which points perpendicular to the propagation direction of the edge states (see Fig.~\ref{fig:UVModel}b). We therefore think of the dipole moment as a pseudospin with up and down states. This interpretation is further supported by the fact that the neutral sector exhibits an emergent 
$\mathfrak{su}(2)_1$ current algebra spanned by operators~\cite{KaneFisherPolchinski1994,ProtopopovGefenMirlin2017}
\begin{equation}
    J^z_{t/b}
    =
    \pm\frac{1}{2\sqrt{2}\pi}
    \partial_x\phi_{t/b,\sigma},
    \qquad
    J_\tau^\pm
    =
    \frac{1}{2\pi a}
    e^{\pm i\sqrt{2}\phi_{\tau\sigma}}.
    \label{eq:bosonicFieldRepOfCurrents}
\end{equation}
We also relate creation and annihilation operators of dipole moments to $x,y$ components via
$J_\tau^\pm=J_\tau^x\pm iJ_\tau^y$
in the usual way.

\subsubsection{Hamiltonian near KFP fixed point}

Relegating hopping and CAR terms to Sec.~\ref{sec:microscopicTheoryOfTheHEterostructure}, it is convenient to express the Hamiltonian~\eqref{eq:UVFullHamiltonian} near the KFP fixed point using charge and neutral modes. The disorder contribution is

\begin{subequations}
\label{eq:FullHamKFPBasis}
\begin{equation}
    H_{\rm dis}
    =
    \frac{1}{a}
    \sum_{\tau=t,b}
    \int\dd x\,
    \left[
        \xi_\tau(x)J_\tau^+
        +
        \xi_\tau^*(x)J_\tau^-
    \right]. \label{eq:DisNearKFP}
\end{equation}
We next discuss the interaction contribution on a phenomenological level. There are interaction terms 
\begin{equation}
    H_\text{kin} + H_{\rm int} = H_\rho + H_\sigma + H_{\rho\sigma}
    \label{eq:Hkin+Hint}
\end{equation}
of three types: First, there are density-density interactions $H_\rho$, which contain intra-edge components absorbed into the mode velocity $v_\rho$ (we assume mirror symmetry with respect to reflections along the edge so that both edges are equivalent) and inter-edge components $g_\rho$. Second, there are dipole-dipole interactions $H_{\sigma} = H^{(0)}_{\sigma} + H_{\sigma}^{tb}$ -- in the most general case, there are 21 different terms connecting any of the three currents of either edge to another current. However, in the vicinity of the KFP fixed point, each edge is SU(2) symmetric so that there is only a single intra-edge component $v_\sigma$ while interedge interactions (denoted $\lambda_i$) may be assumed to be pseudo-spin diagonal. Finally, density-dipole interactions $H_{\rm \rho \sigma}$ vanish by definition at the KFP fixed point. In the vicinity of the KFP fixed point, the interactions described above are marginal and, therefore, serve as natural perturbations to two decoupled edges. Among all operators constructed from four-body interactions between fractional quasiparticles on the edges that do not transfer fractional charge, these are the most relevant interactions. All contributions to Hamiltonian \eqref{eq:Hkin+Hint} can be explicitly written as

\begin{align}
    H_\rho
    &=
    \frac{3}{2}
    \frac{u_\rho}{2\pi}
    \int\dd x\,
    \left[
        \frac{1}{K_\rho}
        (\partial_x\Theta_\rho)^2
        +
        K_\rho
        (\partial_x\Phi_\rho)^2
    \right], \label{eq:Hrho}
    \\
    H^{(0)}_\sigma
    &=
    \frac{2\pi v_\sigma}{3}
    \sum_{\substack{\tau = t,b, \\ i=x,y,z}}
    \int\dd x\,
    J_\tau^iJ_\tau^i, \\
    H_{\sigma}^{tb}&=
    2\pi v_\sigma
    \sum_{i=x,y,z}
    \int\dd x\, 
    \lambda_{i}
    J_t^iJ_b^i, \label{eq:Hsigma} \\
    H_{\rho \sigma} & = \sum_{\substack{\tau, \tau' = t,b, \\ i=x,y,z}} \int \dd x \tilde{g}_{\tau \tau'}^i \partial_x \phi_{\rho,\tau} J_{\tau'}^i. \label{eq:rhosigmaIA}
\end{align}
\label{eq:FullInterTrenchInteraction}
\end{subequations}
The charge-density interaction $g_\rho$ is absorbed into the renormalized velocity $u_\rho = v_\rho\sqrt{1-4\pi^2(g_\rho/v_\rho)^2}$ and the charge Luttinger parameter $ K_\rho = \sqrt{\frac{1-2\pi g_\rho/v_\rho}{1+2\pi g_\rho/v_\rho}}$. The constants $v_\sigma$ and $v_\rho$ denote, respectively, the mode velocities of the neutral and $2e/3$ charge mode at the KFP fixed point at which $\tilde g_{\tau, \tau'}^i$ vanishes by definition \cite{KaneFisherPolchinski1994}. 
The precise values of the couplings $g_\rho$ and $\lambda_i$ depend on the interaction kernel $U$ in Eq.~\eqref{eq:UVFullHamiltonian}.
In the following, $v_\sigma/a$ will serve as the reference energy scale. 

The fields $\Phi_p$ and $\Theta_p$ are nonchiral superpositions of the chiral top and bottom fields:
\begin{subequations}
\label{eq:NonChiral}
\begin{align}
    \Phi_\sigma
    &=
    \frac{1}{\sqrt{2}}
    (\phi_{t\sigma}+\phi_{b\sigma}),
    &
    \Theta_\sigma
    &=
    \frac{1}{\sqrt{2}}
    (\phi_{t\sigma}-\phi_{b\sigma}),
    \\
    \Phi_\rho
    &=
    \frac{1}{\sqrt{6}}
    (\phi_{t\rho}+\phi_{b\rho}),
    &
    \Theta_\rho
    &=
    \frac{1}{\sqrt{6}}
    (\phi_{t\rho}-\phi_{b\rho}).
\end{align}
\end{subequations}
with the corresponding commutations relations 
\begin{subequations}
\begin{align}
    [\Phi_\sigma(x), \Theta_\sigma(x')] & = -i2\pi \theta(x-x'),\\
    [\Phi_\rho(x),\Theta_\rho(x')] &  =i\frac{2\pi}{3}\theta(x-x'). 
\end{align}
\end{subequations}
Analogously to the charge sector, the neutral part of the Hamiltonian can be rewritten as
\begin{equation}
    \begin{split}
        H_\sigma
        &=
        \frac{u_\sigma}{4\pi}
        \int\dd x\,
        \left[
            \frac{1}{K_\sigma}
            (\partial_x\Theta_\sigma)^2
            +
            K_\sigma
            (\partial_x\Phi_\sigma)^2
        \right]
        \\
        &-
        \frac{v_\sigma}{2\pi a^2}
        \int\dd x\,
        \left[
            \lambda_b\cos(2\Theta_\sigma)
            +
            \lambda_p\cos(2\Phi_\sigma)
        \right].
    \end{split}
    \label{eq:sigmaAbelianBosonized}
\end{equation}
Here,
$u_\sigma=v_\sigma\sqrt{1-(\lambda_z/2)^2}$ and $K_\sigma=\sqrt{\frac{1-\lambda_z/2}{1+\lambda_z/2}}$
are the renormalized neutral-mode velocity and Luttinger parameter, respectively. We further define the couplings
\begin{equation}
    \lambda_b
    =
    \frac{\lambda_x+\lambda_y}{2},
    \qquad
    \lambda_p
    =
    \frac{\lambda_x-\lambda_y}{2}. \label{eq:lambdabp}
\end{equation}
Their physical interpretation corresponds to backscattering ($\lambda_b$) and pairing ($\lambda_p$) of the dipole moments \cite{VaeyrynenGefen2022}. Since $\lambda_b$ and $\lambda_p$ couple the $J^\pm_\tau$ operators, we will refer to them 
as exchange interactions in terms of the pseudo-spin. Accordingly, $\lambda_z$ describes the dipole density interactions. 

\subsubsection{Dirty case}
\label{sec:DirtyReview}

To deal with the potentially non-zero disorder, we follow KFP~\cite{KaneFisherPolchinski1994} in that the term $H_\xi$ can be interpreted as a random magnetic field acting in the $x$-$y$ plane of the dipole moment. A suitable local SU(2)
gauge transformation 
can absorb the disorder coupling into the definition of the current operators $J_i$. 

To this end, we briefly review the transformation behavior of the $\mathfrak{su}(2)_1$ fields in the adjoint and fundamental representations.
Under local SO(3) transformations, the current algebra transforms as \cite{ProtopopovGefenMirlin2017, VaeyrynenGefen2022}
\begin{equation}
    J_{t/b}^{i} = S_{t/b}^{ij}\widetilde J_{t/b}^{j} \pm h^i_{t/b},\quad h^i_{\tau}=\frac{1}{8\pi}\epsilon_{ijk}\left[S_\tau\partial_x S_\tau^{\mathrm T}\right]_{jk},
    \label{eq:STrafo}
\end{equation}
where
\begin{equation}
    S_\tau(x)
    =
    \mathcal T_x
    \exp\left(
        \int_{- \infty}^{x}dx'\,
        M_\tau(x')
    \right), \quad S_\tau (- \infty)=\mathbf 1.
\end{equation}
Here, $\mathcal T_x$ denotes path ordering, and
\begin{equation}
    M_\tau^{ij}(x)
    =
    4\pi\epsilon_{ijk}h_\tau^k.
\end{equation}
The matrix $M_\tau$ can further be expanded as
\begin{equation}
    M_\tau
    =
    \sum_{i=x,y,z}
    m_\tau^iL_i,
    \qquad
    m_\tau^i
    =
    4\pi h_\tau^i, \label{eq:littlem}
\end{equation}
where $L_i$ are the generators of SO(3) (adjoint representation). The current operators $\widetilde J_{\tau}^i$ have the same representation in terms of bosonic fields as in \eqref{eq:bosonicFieldRepOfCurrents} but with the transformed bosonic fields $\tilde \phi_{\tau\sigma}$.

The $\mathfrak{su}(2)_1$ current algebra also has a spinor representation given by the vertex operators 
\begin{equation}
    \mathbf z_\tau = \begin{pmatrix}
        z_{\tau+} \\
        z_{\tau -}
    \end{pmatrix},\quad\text{where}\quad z_{\tau\sigma} = e^{\pm i\phi_{\tau\sigma}/\sqrt{2}}.
    \label{eq:SpinorRep}
\end{equation}
The spinor $\mathbf z_\tau$ transforms according to the fundamental representation of SU(2) corresponding to the orthogonal rotation $S_\tau(x)$, i.e.
\begin{equation}
   \mathbf z_\tau = U_{\tau}(x) \tilde{\mathbf z}_\tau,
\end{equation}
where 
\begin{equation}
    U_\tau = \mathcal T_x \exp\left(-\frac{i}{2}\int_{- \infty}^x\dd{x'} m_\tau ^i \sigma_i\right)
\end{equation}
with $\sigma_i$ being the Pauli matrices in $\pm$ space and $m_\tau^i$ is defined in Eq.~\eqref{eq:littlem}.

We return to the disorder potential $\xi_\tau(x)$ entering Eq.~\eqref{eq:DisNearKFP} which may be absorbed into ``comoving'' fields by choosing \cite{VaeyrynenGefen2022}
\begin{equation}
    h_\tau^i =
    \begin{cases}
        s_\tau\dfrac{\frac{2}{3}\xi_\tau^i -\lambda_i\xi_{\bar\tau}^i} {v_\sigma\left(\lambda_i^2-(2/3)^2\right)}, & \text{for } i=x,y, \\[6pt] 0, & \text{for } i=z,
    \end{cases}
    \label{eq:hValuesForGeneralDisorder}
\end{equation}
where $\bar\tau$ denotes the edge opposite to $\tau$. 

The matrix $U_\tau(x)\in\text{SU(2)}$, or equivalently its adjoint representation $S_\tau(x)\in \text{SO}(3)$, may be viewed as the endpoint of a random walk on the corresponding group which is induced from a random walk $h_\tau^i(x)$ in the tangent space by means of the path-ordered exponential map. 
The spatial coordinate parametrizes the walk with boundary condition $U_\tau(x)\vert_{x \rightarrow - \infty}=\mathbf 1$. For short-range disorder, the probability measure $\mathcal P(U_\tau) \mathcal D{U_\tau}$, which inherits the white-noise lack of spatial correlations on distances larger than the disorder length scale $l_\text{dis} \sim {v_\sigma^2}/{W}$, approaches the normalized Haar measure $\mathcal D {U_\tau}$,  
that is $\mathcal P(U_\tau) \mathcal D {U_\tau}{\simeq}\mathcal D{U_\tau}$, see App. \ref{AppSec:disorderGaugeTrafo} for details.

\subsection{Tunneling and CAR near the KFP Fixed Point}
\label{sec:microscopicTheoryOfTheHEterostructure}

Having rewritten all interaction terms in a basis convenient near
the KFP fixed point, we now turn to tunneling and pairing terms in the Hamiltonian \eqref{eq:UVFullHamiltonian}.

\subsubsection{Electron Operator}
\label{sec:Elop}

At the KFP fixed point, each edge possesses two most relevant electron operators,
\begin{equation}
    \psi_{\tau\pm}
    =
    \frac{\kappa_\pm}{\sqrt{2\pi a}}
    e^{
        i\sqrt{\frac{3}{2}}\phi_{\tau\rho}
        \pm
        i\sqrt{\frac{1}{2}}\phi_{\tau\sigma}
    },
    \label{eq:PMfermionDefinition}
\end{equation}
with identical scaling dimensions $\Delta_\pm=1$. The auxiliary Majorana fermions $\kappa_\pm$ ensure the correct anticommutation relations, and the physical Hilbert space is fixed by the constraint $i\kappa_+\kappa_-=1$. In the original field basis, the operator $\psi_{\tau-}\sim e^{i\phi_{\tau1}}$ corresponds to an original electron, while the $\psi_{\tau+}$ electron operator is the combination of 2 electrons and 3 quasi-holes. 

In the important case when the KFP fixed point is disorder stabilized, it is a crucial observation that the quoted most relevant electron operators presented in Eq.~\eqref{eq:PMfermionDefinition} are not written in the laboratory frame but rather in the comoving frame associated to Eq.~\eqref{eq:hValuesForGeneralDisorder}. Specifically, it is convenient to use Eq.~\eqref{eq:SpinorRep} and collect the two electron operators into the spinor $\Psi_{\tau} = (\psi_{\tau+}, \psi_{\tau-})$ that can be rewritten as
\begin{equation}
    \Psi_\tau = \frac{1}{\sqrt{2\pi a}}e^{i\sqrt{\frac{3}{2}}\phi_{\tau\rho}} D_\kappa \mathbf z_\tau,\text{ where } D_\kappa = \begin{pmatrix}
        \kappa_+ & 0 \\ 0 & \kappa_-
    \end{pmatrix}. \label{eq:SpinorElectron}
\end{equation}
This demonstrates that a rotation $S_\tau(x)$ in the current algebra also causes an according rotation of the electron spinor $\Psi\sim \mathbf z$, that is
\begin{equation}
    \Psi_\tau = \hat{ U}_\tau \tilde{\Psi}_\tau,\text{ with } \hat{U}_\tau = D_\kappa U_\tau D_\kappa. \label{eq:PsiSpinor}
\end{equation}
The physical electron entering Eq.~\eqref{eq:UVFullHamiltonian} is then given by the superposition
\begin{equation}
    \psi_\tau
    =
    c_{\tau+}\psi_{\tau+}
    +
    c_{\tau-}\psi_{\tau-},
    \label{eq:electronOperator}
\end{equation}
where the coefficients satisfy
$|c_{\tau+}|^2+|c_{\tau-}|^2=1$
and are determined by the microscopic details of the edges. Note that generically, these coefficients can also have a spatial dependence.

\subsubsection{Pairing and Tunneling terms}

In the following, we express the tunneling ($H_t$) and pairing ($H_\Delta$) Hamiltonians at the KFP fixed point in terms of the fermionic operators $\psi_{\tau\pm}$.  
At the KFP fixed point without inter-edge interactions, the electron operator is given by Eq.~\eqref{eq:electronOperator} and has scaling dimension $\Delta_e=1$. Thus, pairing and tunneling are marginal perturbations around the KFP fixed point.

Using the spinor representation~\eqref{eq:SpinorElectron}
one can write hopping and CAR terms of Eq.~\eqref{eq:UVFullHamiltonian} as
\begin{subequations}
\begin{align}
    H_t & 
   =
    -\int \dd x \Psi_t^\dagger\mathcal T\Psi_b+\text{H.c.}, \label{eq:SpinorHopt}\\
   H_\Delta &=
 \int d x   \Psi_t^{\mathrm T}\mathcal D\Psi_b + \text{H.c.},
\end{align}\label{eq:SpinorHopDelta}
\end{subequations}
in terms of
the tunneling matrix
\begin{equation}
    \mathcal T
    =
    \sum_{\alpha=0,x,y,z}
    t_\alpha\sigma_\alpha,
\end{equation}
and the matrix-valued proximity induced order parameter 
\begin{equation}
    \mathcal D
    =
    (d_0+\mathbf d\cdot\boldsymbol{\sigma})
    i\sigma_y.
\end{equation}
 The singlet pairing amplitude $d_0$ is chosen to be real, while the triplet pairing amplitudes $d_i$, with $i\in\{x,y,z\}$, are complex numbers with phases $\delta_i$. These amplitudes are collected into a vector
\begin{equation}
    \mathbf d
    =
    (d_x,d_y,d_z).
\end{equation}
For the disordered case, it is worthwhile to highlight that Eq.~\eqref{eq:SpinorHopDelta} is written in the lab frame and that both $\mathcal T$ and $\mathcal D$ may in the most generic case be spatially dependent. Their spatial dependence absorbs the spatial dependence of $t,\Delta$ discussed above Eq.~\eqref{eq:DisorderPotential} and of $c_{\tau,\pm}$ discussed below Eq.~\eqref{eq:electronOperator}. 
The tunneling strength $t$ and pairing amplitude $\Delta$ are determined by the overlap of the exponentially localized top and bottom electron states.
They can be expected to be exponentially small in the trench width $d$, i.e.,
\begin{equation}
    t,\Delta\sim e^{-d/\zeta_{\text{eff}}},
    \label{eq:tDeltaSmall}
\end{equation}
where $\zeta_\text{eff}$ is the effective localization length scale describing the wave-function envelope perpendicular to the edge.

Using Eqs.~\eqref{eq:NonChiral} and \eqref{eq:PMfermionDefinition} to  individually bosonize each channel (see App.~\ref{secApp:Bosonization}), yields 
\begin{widetext}
\begin{subequations}
\begin{align}
        H_t
        &=
         \int_{-\infty}^{0}\dd x\,\frac{2t_0}{\pi a}
        \cos(\Theta_\sigma)
        \sin(3\Theta_\rho)
        +
        \frac{2}{\pi a}
        \int_{-\infty}^{0}\dd x\,
        \left[
            t_z\sin(\Theta_\sigma)
            -
            t_x\sin(\Phi_\sigma)
            -
            t_y\cos(\Phi_\sigma)
        \right]
        \cos(3\Theta_\rho),
    \label{eq:BosonizedHt}\\
        H_\Delta
        &= 
        -\int_{0}^{\infty}\dd x\, \frac{2d_0}{\pi a}
        \cos(\Theta_\sigma)
        \sin(3\Phi_\rho)-
        \int_{0}^{\infty}\dd x\,
        \frac{2|d_x|}{\pi a}
        \sin(\Phi_\sigma)
        \cos(3\Phi_\rho+\delta_x)
        \notag \\
        &+
        \int_{0}^{\infty}\dd x\,
        \frac{2|d_y|}{\pi a}
        \cos(\Phi_\sigma)
        \cos(3\Phi_\rho+\delta_y)
        +
        \int_{0}^{\infty}\dd x\,
        \frac{2|d_z|}{\pi a}
        \sin(\Theta_\sigma)
        \cos(3\Phi_\rho+\delta_z).
    \label{eq:BosonizedHDelta}
\end{align}
\label{eq:tDeltaBosonized}
\end{subequations}
\end{widetext}

\section{Clean Case}
\label{sec:cleanCase}

\begin{figure*}
    \centering
    \includegraphics[width=\linewidth]{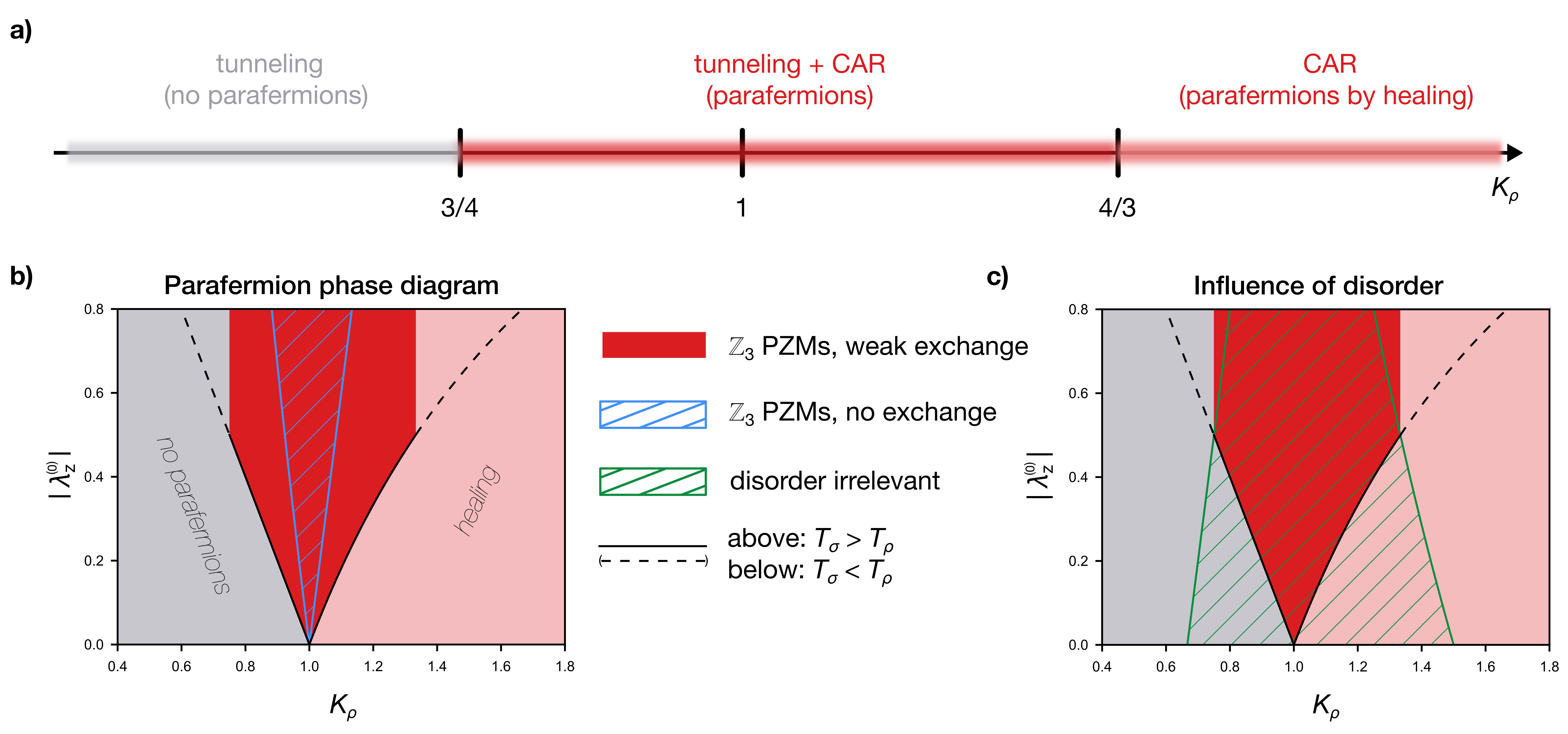}
    \caption{\textbf{a)} 
    Phase diagram for strong dipole exchange interactions, cf.~Sec.~\ref{sec:StrongExchange} where the neutral sector gaps out first ($T_\sigma > T_\rho$). In the red central region, tunneling and CAR are both relevant perturbations to the low-energy theory leading to PZMs at the interface between tunneling and CAR regions. For $K_\rho>4/3$, only CAR processes are RG relevant. However, 
    healing (see Sec.~\ref{sec:MinimaAndDerivationOfParafermions}) may nevertheless lead to PZMs. \textbf{b)} Parafermion phase diagram for weak exchange interaction, cf.~Sec.~\ref{sec:WeakExchange}, in the parameter space of the Luttinger parameter $K_\rho$ and the bare dipole coupling $|\lambda_z^{(0)}|$: In the dark red region, the low-energy theory allows for PZMs at the interface of a tunneling and CAR region. Above the black (dashed) line, the exchange coupling is the most relevant, i.e., $|\lambda_z^{(0)}|>y_t,y_d$ and $T_\sigma > T_\rho$. The blue hatched region denotes the region in which a parafermion can exist when gapping out neutral and charge sectors simultaneously in the absence of any dipole exchange interactions. \textbf{b)} Influence of the disorder terms on the parafermionic phase: The green hatched region shows where all effective disorder strengths $\mathscr D_{\Phi,\Theta}, \mathscr C_{\Phi,\Theta}$ are irrelevant.}
    \label{fig:PhaseDiagram}
\end{figure*}

The central problem of the spin-polarized $\nu=2/3$ edge theory is that the tunneling and pairing Hamiltonians are described by a complicated  
field-theory, which is generically not SU(2) invariant, and in which different terms compete to pin different sectors of the neutral modes, see Eqs. \eqref{eq:BosonizedHt} and \eqref{eq:BosonizedHDelta}. In combination with marginal inter-edge interactions, however, the renormalization group (RG) flow can favor certain tunneling and pairing channels over others.
To see how inter-edge interactions can give rise to a double sine-Gordon model and, consequently, to parafermions at the interface between tunneling and pairing, we first consider the clean case ($\xi_\tau=0$) in which fine-tuned intra-edge interactions drive the system towards the KFP fixed point. This allows us to demonstrate the main mechanism in a simpler setting before turning to the more generic case in the next section.

\subsection{Renormalization Group Flow and Interaction Regimes}

Assuming that the interactions in Hamiltonian \eqref{eq:UVFullHamiltonian} are predominantly electrostatic, we expect the charge interaction $g_\rho$ to dominate over the dipole interactions, such that $4\pi g_\rho/v_\sigma \gg \lambda_{x,y,z}$. We therefore treat the latter perturbatively. Furthermore, the tunneling and CAR amplitudes are exponentially suppressed with increasing trench width $d$ and are assumed to be parametrically small compared with all other energy scales. Consequently, we neglect their feedback on the remaining couplings in the subsequent RG analysis.

\subsubsection{Renormalization Group Flow Equations}

To the first-loop order, the dipole interaction couplings renormalize according to the well-known flow equations of anisotropic ${\mathfrak{su}}(2)_1$ models
\begin{align}
    \frac{d\lambda_x}{d\ell} = \lambda_y\lambda_z,
    \,\,\,\,\frac{d\lambda_y}{d\ell} = \lambda_z\lambda_y,\,\,\,\,
    \frac{d\lambda_z}{d\ell} = \lambda_x\lambda_y.
    \label{eq:RGFlowDipol}
\end{align}
In the presence of inter-edge interactions, the tree-level flow equations for the   dimensionless tunneling couplings $\mathscr t_{\alpha}=8t_\alpha a/v_\sigma$ are (see App. \ref{AppSec:DerivationRGFlowEquations}) 
\begin{subequations}
\begin{align}
    \frac{d\mathscr t_{0}}{d\ell}
    &=
    \left(
    \frac{3}{2}
    -\frac{3}{2}K_\rho
    +\frac{\lambda_x +\lambda_y+\lambda_z}{4}
    \right)
    \mathscr t_{0},\label{eq:RGt0}\\ 
    \frac{d\mathscr t_{x}}{d\ell}
    &=
    \left(
    \frac{3}{2}
    -\frac{3}{2}K_\rho
    +\frac{-\lambda_x + \lambda_y -\lambda_z}{4}
    \right)
    \mathscr t_{x}, \label{eq:RGFlowTx}\\
    \frac{d\mathscr t_{y}}{d\ell}
    &=
    \left(
    \frac{3}{2}
    -\frac{3}{2}K_\rho
    +\frac{\lambda_x-\lambda_y-\lambda_z}{4}
    \right)
    \mathscr t_{y},\label{eq:RGty}\\
    \frac{d\mathscr t_{z}}{d\ell}
    &=
    \left(
    \frac{3}{2}
    -\frac{3}{2}K_\rho
    +\frac{-\lambda_x-\lambda_y+\lambda_z}{4}
    \right)
    \mathscr t_{z}.\label{eq:RGtz}
\end{align}
\label{eq:tDeltaFlow}
\end{subequations}
The flow equations for the corresponding dimensionless CAR amplitudes $\mathscr d_\alpha = 8d_\alpha a/v_\sigma$ are identical, but one has to replace $K_\rho \rightarrow 1/K_\rho$. 

Since  $\mathscr t_\alpha$ and $\mathscr d_\alpha$ are parametrically small compared to the density interaction $4\pi g_\rho/v_\rho$, their RG corrections to $K_\rho$ by tunneling and CAR processes will be weak. Since the Hamiltonian \eqref{eq:Hkin+Hint} (at the KFP fixed point) also does not contain any term that renormalizes the charge density interaction, we treat $K_\rho$ as a microscopic parameter that does not flow, that is 
\begin{equation}
    \frac{d K_\rho}{d \ell} =0.
\end{equation}

\subsubsection{Classification of Fixed Points}

Since the couplings $\mathscr t_\alpha$ and $\mathscr d_\alpha$ only follow the RG flow of the dipole interactions but do not influence the $\lambda_{x,y,z}$ couplings, the IR physics is completely determined by the fixed points of the flow equations~\eqref{eq:RGFlowDipol}.

\begin{figure}
    \centering
    \includegraphics[width=\linewidth]{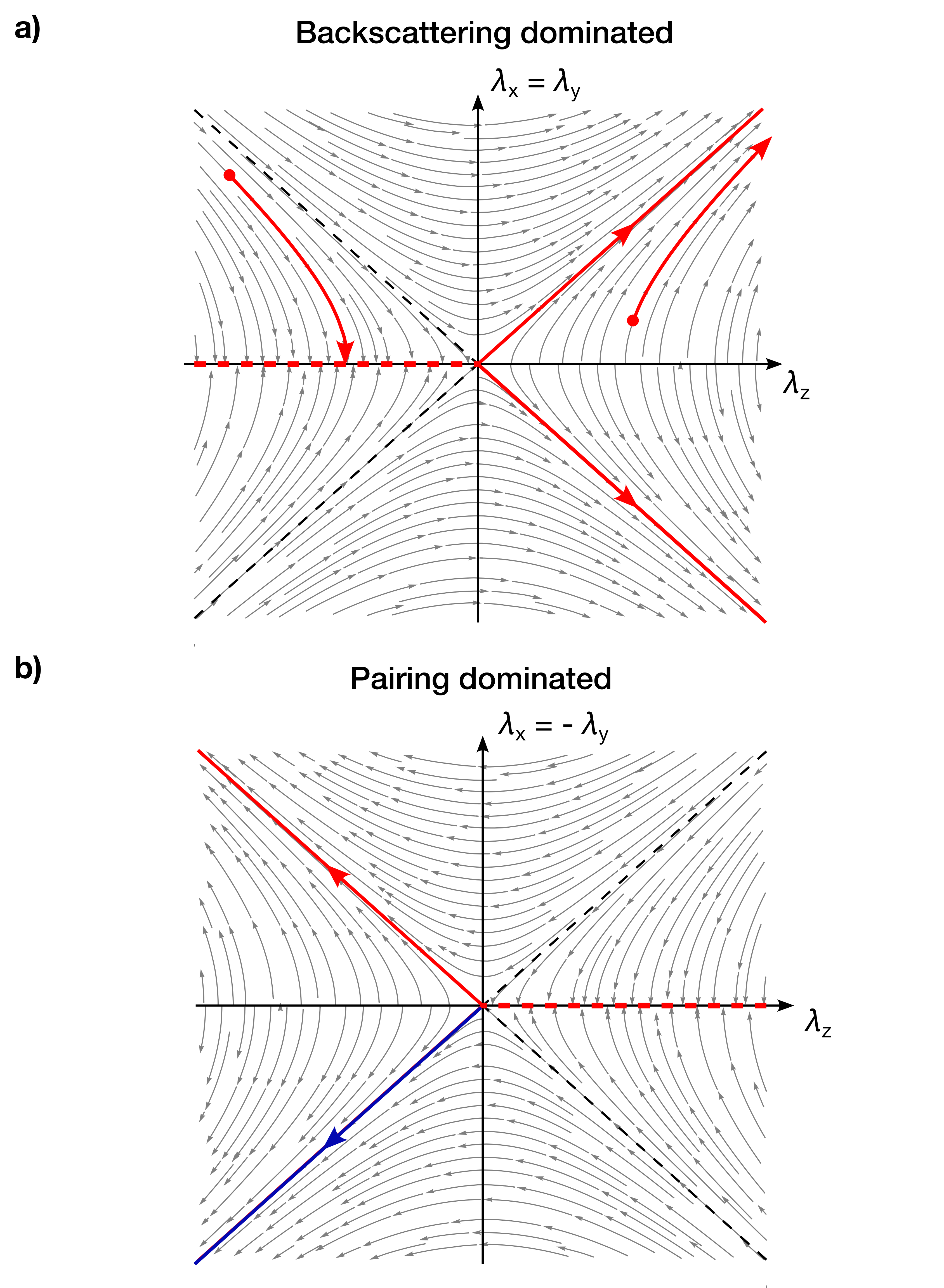}
    \caption{Flow diagrams of the dipolar interaction couplings $\lambda_x$, $\lambda_y$, and $\lambda_z$ along the line cuts $\lambda_x=\lambda_y$ (\textbf{a}) and $\lambda_x=-\lambda_y$ (\textbf{b}). The dashed gray lines denote the separatrices between the strong-coupling and Gaussian regimes. The red dashed lines indicate stable Gaussian fixed points. The red arrows indicate the strong-coupling rays, whereas the blue arrow marks the disorder-invariant strong-coupling ray.}
    \label{fig:FlowDiagrams}
\end{figure}

In the following, we classify all IR cases of the dipole interaction \cite{VaeyrynenGefen2022}.
Qualitatively, the system may end up in two different situations. First, one starts/ends up on one of the three fixed point axes
\begin{equation}
   \boldsymbol{\lambda} \equiv (\lambda_x,\lambda_y,\lambda_z) \longrightarrow \lambda \begin{cases}
     (1, 0, 0) \\
     (0, 1, 0) \\
     (0, 0, 1)
    \end{cases},
\end{equation}
where $\lambda\in \mathbb R$. In this case, we can always find a global rotation of the dipole moments that rotates the interaction to a pure $J_t^z J_b^z \sim \partial_x\phi_t \partial_x\phi_b$ interaction and, thus, to a Gaussian theory.  

Second, the system can develop a gap in the neutral sector by approaching one of the four strong coupling rays 
\begin{equation}
    \boldsymbol{\lambda}  
    \longrightarrow  \begin{cases}
     \lambda(\pm1, \pm1, 1) & \lambda>0 \\
     \lambda(\pm1, \mp 1, 1) & \lambda <0
    \end{cases}.
\end{equation}

This behavior can be understood best when we consider the two special cases of dominant dipole backscattering $|\lambda_b| \gg |\lambda_p|$ ($\lambda_x\approx \lambda_y$) and dominant dipole pairing $|\lambda_b| \ll |\lambda_p|$ ($\lambda_x\approx -\lambda_y$) as shown in Fig.~\ref{fig:FlowDiagrams}a) and Fig.~\ref{fig:FlowDiagrams}b), respectively.
For backscattering dominating the physics, the system approaches a Gaussian theory with a renormalized Luttinger parameter $K^*_\sigma>1$, as long as the bare $\lambda^{(0)}_z < -|\lambda_b|$. For $\lambda^{(0)}_z > -|\lambda_b^{(0)}|$ and $\lambda^{(0)}_b>0$ the system flows towards the isotropic strong coupling ray with $\lambda(1,1,1)$ with $\lambda > 0$. On the other hand, if $\lambda^{(0)}_z > -|\lambda_b^{(0)}|$ and $\lambda^{(0)}_b<0$ the system approaches $\boldsymbol{\lambda} \rightarrow \lambda(-1, -1, 1)$ with $\lambda >0$.

In the opposite case, where the dipole pairing determines the physics, a Gaussian theory will be reached for $\lambda^{(0)}_z>|\lambda_p^{(0)}|$ with $K^*_\sigma <1$.  For $\lambda^{(0)}_z<|\lambda_p^{(0)}|$ the system approaches the strong coupling ray $\boldsymbol{\lambda} \rightarrow \lambda(1, -1, -1)$ with $\lambda >0$ when $\lambda^{(0)}_p > 0$ and $\boldsymbol{\lambda} \rightarrow \lambda(-1, 1, -1)$ for $\lambda^{(0)}_p < 0$.

\subsection{Different Scenarios and Pinning Mechanism}

As will be demonstrated in this section, gapping out the charge modes $\Phi_\rho$ and $\Theta_\rho$ is the crucial step in realizing parafermions.  
As apparent from Eqs.~\eqref{eq:BosonizedHt} and \eqref{eq:BosonizedHDelta}, a general theory that takes all tunneling and CAR channels into account suffers from the problem of competing tendencies to pin the neutral sector as well. However, inter-edge dipole interactions serve as a tool to favor certain tunneling and CAR channels.

However, we have to be careful about the mechanism that pins the neutral field, as either the dipole exchange interaction can pin $\Theta_\sigma$ or $\Phi_\sigma$ when flowing to a strong coupling ray, or the tunneling and CAR terms pin the neutral field simultaneously with the charge fields. To distinguish these two different situations, we define the temperature scales $T_\sigma$ and $T_\rho$, where, respectively, the dipole exchange couplings $\lambda_{x,y}$ and the tunneling/CAR $\mathscr t_\alpha/\mathscr d_\alpha$ become order one. 

In the following, we consider three different scenarios with slightly different mechanisms of how parafermions can emerge in the effective low energy theory. 
First we consider the case of \textit{strong exchange coupling}, where $\lambda_{x,y}\sim \lambda_z \gg \mathscr t_\alpha, \mathscr d_\alpha$. Here, it is effectively guaranteed that the exchange coupling will reach $\mathcal O(1)$ first, i.e. $T_\sigma >T_\rho$, and the neutral field gapping is due to dipole exchange interactions. The second case corresponds to \textit{weak exchange couplings}, $\lambda_z \gg \lambda_{x,y}\sim  \mathscr t_\alpha, \mathscr d_\alpha$. Here, whether $T_\sigma > T_\rho$ or $T_\sigma < T_\rho$ depends on the parameter space $(K_\rho, \lambda_z^{(0)})$. Last, we consider the case where exchange couplings are absent, and $T_\sigma$ is ill-defined. 

\subsubsection{Strong Exchange Coupling}
\label{sec:StrongExchange}

For comparable exchange couplings, $\lambda_{x,y}\sim\lambda_z$, and parametrically weak tunneling and pairing amplitudes, we expect exchange interactions to provide the dominant mechanism for gapping the neutral sector, with $T_\sigma>T_\rho$. Depending on the bare couplings $\lambda_i^{(0)}$, the dipole interactions may flow toward one of the four strong-coupling rays shown in Fig.~\ref{fig:FlowDiagrams}. Each ray preferentially enhances one of the four tunneling/CAR channels in the RG equations
\begin{equation}
    \frac{d\mathscr t_{\alpha}}{d\ell}
    = \left(\frac{3}{2}-\frac{3}{2}K_\rho+\frac{\mathscr{n}_\alpha}{4}\lambda\right)\mathscr t_{\alpha}. \label{eq:nRG}
\end{equation}
As before, RG equations for $\mathscr d_\alpha$ follow $K_\rho \rightarrow 1/K_\rho$. The values of $\mathscr{n}_\alpha$ depend on the specific strong-coupling ray and are summarized in Tab.~\ref{tab:StrongCouplingRays}.
For example, along the isotropic ray $\boldsymbol{\lambda} \rightarrow \lambda(1,1,1)$, 
exchange interactions therefore, enhance
$\mathscr t_0$ relative to the other channels, favoring it as the dominant perturbation governing the subsequent low-energy behavior. The other three strong-coupling rays exhibit analogous behavior, each favoring a different channel. The exchange-induced corrections to the flow equations along all four rays are summarized in Tab.~\ref{tab:StrongCouplingRays}.

\begin{table}[htbp]
    \centering
    \renewcommand{\arraystretch}{1.2}
    \setlength{\tabcolsep}{12pt}
    \begin{tabular}{@{}lrrrr@{}}
        \toprule
        & \multicolumn{4}{c}{Values of $\mathscr{n}_\alpha$ in Eq.~\eqref{eq:nRG}}\\
        \cmidrule(l){2-5}
        Strong-coupling ray & $0$ & $x$ & $y$ & $z$ \\
        \midrule
        $\boldsymbol{\lambda} \rightarrow \lambda( 1,  1,  1)$ & $ 3$ & $-1$ & $-1$ & $-1$ \\
        $\boldsymbol{\lambda} \rightarrow \lambda(-1,  1, -1)$ & $-1$ & $ 3$ & $-1$ & $-1$ \\
        $\boldsymbol{\lambda} \rightarrow \lambda( 1, -1, -1)$ & $-1$ & $-1$ & $ 3$ & $-1$ \\
         $\boldsymbol{\lambda} \rightarrow \lambda(-1, -1,  1)$ & $-1$ & $-1$ & $-1$ & $ 3$ \\
        \bottomrule
    \end{tabular}
    \caption{Coefficients entering Eq.~\eqref{eq:nRG} associated with the four strong-coupling rays.
             All entries in columns $0$, $x$, $y$, and $z$
             are multiplied by $\lambda/4$.}
    \label{tab:StrongCouplingRays}
\end{table}

Along the isotropic strong-coupling ray, where $\lambda_p=0$ and $\lambda_b=\lambda$, the neutral-sector Hamiltonian $H_\sigma$ semiclassically pins $\Theta_\sigma$ to one of the two minima, 0 or $\pi$. This occurs at the RG scale $\ell^*$, where $\lambda(\ell^*)\sim\mathcal O(1)$, and generates a neutral gap 
$m_\sigma\sim T_\text{KFP} \exp\left(-\frac{c}{|\lambda_b^{(0)}|}\right)$, where $c$ is an order-one coefficient determined by the ratio $\lambda_z^{(0)}/|\lambda_b^{(0)}|$.

In a chosen pinned vacuum, we write $\Theta_\sigma=\langle\Theta_\sigma\rangle+\vartheta$, where $\vartheta$ describes massive fluctuations about the selected minimum with $\langle\vartheta^2\rangle\sim-\ln(m_\sigma/T_{\mathrm{KFP}})$. Integrating out the gapped neutral sector, therefore, yields the effective low-energy Hamiltonian
\begin{equation}
H_{\mathrm{eff}}
=t_{\mathrm{eff}}\int_{-\infty}^{0}\dd x\sin(3\Theta_\rho)
-d_{\mathrm{eff}}\int_{0}^{\infty}\dd x\sin(3\Phi_\rho),
\label{eq:effectiveDoubleSineGordon}
\end{equation}
where
$ t_{\mathrm{eff}}
=\frac{2}{\pi a}t_0(\ell^*)
e^{i\langle\Theta_\sigma\rangle-\frac{1}{2}\langle\vartheta^2\rangle},
$ and $ d_{\mathrm{eff}}
=\frac{2}{\pi a}d_0(\ell^*)
e^{i\langle\Theta_\sigma\rangle-\frac{1}{2}\langle\vartheta^2\rangle}.$ 
This Hamiltonian has the same double sine-Gordon structure as Eq.~\eqref{eq:reviewBosonizedTunnelingPairing}, which supports $\mathbb Z_3$ PZMs at interfaces between tunneling- and CAR-gapped regions.

At energy scales below the neutral gap, the dimensionless effective couplings obey
\begin{subequations}
\begin{align}
\frac{d\mathscr t_{\mathrm{eff}}}{d\ell}
&=
\left(2-\frac{3}{2}K_\rho\right)
\mathscr t_{\mathrm{eff}},\\
\frac{d\mathscr d_{\mathrm{eff}}}{d\ell}
&=
\left(2-\frac{3}{2K_\rho}\right)
\mathscr d_{\mathrm{eff}}.
\end{align}
\end{subequations}
Both perturbations are relevant for $3/4<K_\rho<4/3$, yielding the phase diagram shown in Fig.~\ref{fig:PhaseDiagram}a). As the dimensionless couplings reach order unity, tunneling pins $\Theta_\rho$ for $x<0$, while CAR pins $\Phi_\rho$ for $x>0$. The charge sector thus becomes gapped on both sides, and its interface supports PZMs. This second stage of gap formation is illustrated by the transition from Fig.~\ref{fig:Mechanisim}c) to Fig.~\ref{fig:Mechanisim}d).

\subsubsection{Weak Exchange Coupling}
\label{sec:WeakExchange}

When the exchange couplings are much weaker than the dipole density interaction, $|\lambda_{x,y}|\ll|\lambda_z|$, their direct contributions to the tunneling and CAR flows can initially be neglected. The sign of the bare $\lambda_z^{(0)}$ then determines which channels are preferentially enhanced: positive $\lambda_z^{(0)}$ favors the 0- and $z$-channels, whereas negative $\lambda_z^{(0)}$ favors the $x$- and $y$-channels. The corresponding accessible strong-coupling rays are of dipole-backscattering and dipole-pairing type, respectively.

Which mechanism reaches strong coupling first, exchange interactions or tunneling/CAR, depends on their bare amplitudes and RG growth rates. To estimate the corresponding scales, we assume that the bare exchange, tunneling, and CAR couplings are comparable, e.g., exponentially small in the trench width $d$. Since these couplings are initially much smaller than $|\lambda_z^{(0)}|$, we neglect their feedback on $\lambda_z$ over most of the weak-coupling flow and approximate $\lambda_z(\ell)\approx\lambda_z^{(0)}$. The relevant flow equations then take the common form
\begin{subequations}
\begin{align}
\frac{d\lambda_\text{ex}}{d\ell}
&=|\lambda_z^{(0)}|\lambda_\text{ex},\\
\frac{d\mathscr t_\alpha}{d\ell}\label{eq:simplefiedRGflowt}
&=y_t\mathscr t_\alpha,\\
\frac{d\mathscr d_\alpha}{d\ell}\label{eq:simplefiedRGflowDelta}
&=y_d\mathscr d_\alpha,
\end{align}
\label{eq>0}
\end{subequations}
where 
\begin{equation}
    \lambda_\text{ex}=\begin{cases}
        \lambda_b, & \lambda_z^{(0)}>0 \\
        \lambda_p, & \lambda_z^{(0)}<0
    \end{cases}, \quad \alpha \in \begin{cases}
        \{0, z\}, & \lambda_z^{(0)}>0 \\
        \{x, y\}, & \lambda_z^{(0)}<0
    \end{cases}
\end{equation}
and
\begin{equation}
y_t=\frac{3}{2}-\frac{3}{2}K_\rho+\frac{|\lambda_z^{(0)}|}{4},
\qquad
y_d=\frac{3}{2}-\frac{3}{2K_\rho}+\frac{|\lambda_z^{(0)}|}{4}.
\end{equation}
Within this approximation, the couplings evolve as
\begin{subequations}
\begin{align}
\lambda_\text{ex}(\ell) &=\lambda_\text{ex}^{(0)}e^{|\lambda_z^{(0)}|\ell},
 \\
\mathscr t_\alpha(\ell)&=\mathscr t_\alpha^{(0)}e^{y_t\ell},
\\
\mathscr d_\alpha(\ell)&=\mathscr d_\alpha^{(0)}e^{y_d\ell}.
\end{align}
\end{subequations}
For comparable bare amplitudes of the competing couplings, their growth rates therefore provide an approximate criterion for which mechanism reaches strong coupling first. In particular,
\begin{equation}
|\lambda_z^{(0)}|>y_t
\quad\text{and}\quad
|\lambda_z^{(0)}|>y_d
\end{equation}
identifies the regime in which exchange interactions gap the neutral sector before tunneling or CAR reaches strong coupling, yielding $T_\sigma>T_\rho$. The (dashed) black lines in Figs.~\ref{fig:PhaseDiagram}b) and \ref{fig:PhaseDiagram}c), defined by $|\lambda_z^{(0)}|=y_t$ and $|\lambda_z^{(0)}|=y_d$, mark the estimated boundaries of this regime. The gap scales \cite{VaeyrynenGefen2022}
 as $m_\sigma \sim T_\text{KFP} \left(2\frac{|\lambda_z^{(0)}|}{|\lambda_\text{ex}^{(0)}|}\right)^{-1/|\lambda_z^{(0)}|}$.

For $\lambda_z^{(0)}>0$, the sign of $\lambda_b$ further selects which of the two favored channels survives below the neutral gap. If $\lambda_b>0$, the flow approaches the isotropic strong-coupling ray, and $H_\sigma$ semiclassically pins $\Theta_\sigma$ to 0 or $\pi$. In either pinned vacuum, the neutral factor of the $0$-channel, ($\cos\Theta_\sigma$), has a nonzero expectation value, whereas that of the $z$-channel, ($\sin\Theta_\sigma$), vanishes. Integrating out the neutral sector thus retains the 0-channel with a renormalized amplitude and eliminates the $z$-channel contribution to first order in its coupling. For $\lambda_b<0$, the minima shift to $\Theta_\sigma=\pi/2$ or $3\pi/2$, reversing the roles of the two channels.

The resulting phase diagram in the $(K_\rho,|\lambda_z^{(0)}|)$ plane is shown in Fig.~\ref{fig:PhaseDiagram}b). The red region identifies where we expect parafermion zero modes at interfaces between tunneling- and CAR-gapped regions. Within the mechanism considered here, this requires both exchange-driven neutral pinning, as estimated by the inequalities above, and $3/4<K_\rho<4/3$, so that the surviving tunneling and CAR perturbations are both relevant below the neutral gap.

The same reasoning applies for $\lambda_z^{(0)}<0$, where dipole pairing gaps the neutral sector. The estimated phase diagram has the same shape, but neutral pinning now selects the $x$- or $y$-channel to govern the subsequent charge-sector gap formation.

\subsubsection{Absent
exchange coupling and stability beyond KFP fixed point}

Finally, we consider the realization of PZMs in the absence of exchange interactions, $\lambda_{x,y}=0$. In this limit, the RG equations reduce to Eqs.~\eqref{eq:simplefiedRGflowt} and \eqref{eq:simplefiedRGflowDelta}. Although exchange interactions no longer select a single tunneling/CAR channel, the dipole density interaction still distinguishes between two pairs: $\lambda_z>0$ favors the $0$- and $z$-channels, whereas $\lambda_z<0$ favors the $x$- and $y$-channels and, thereby, gets rid of the competition in fluctuations between the two dual fields $\Phi_\sigma$ and $\Theta_\sigma$. This partial channel selection is already sufficient to allow PZMs without exchange interactions, provided that tunneling and CAR are both relevant, $y_t,y_d>0$. For more details, see App.~\ref{secApp:parafermionsVanila}.

Since the neutral sector is not gapped independently by exchange interactions, these relevance conditions depend on both $K_\rho$ and $\lambda_z$. The resulting region supporting this mechanism is shown by the blue hatched area in Fig.~\ref{fig:PhaseDiagram}b). It is substantially smaller than the region accessible when exchange interactions first gap the neutral sector, after which the relevance of the effective tunneling and CAR perturbations is controlled solely by the charge Luttinger parameter.

\section{Dirty Case}
\label{sec:dirtyCase}

The previous section relied on an approximately fine-tuned interaction: The coupling constants $\tilde g_{\tau \tau'}^i$ of Eq.~\eqref{eq:rhosigmaIA} were assumed to be small. We now turn to the more generic case in which quenched disorder with $\xi_\tau(x)\neq0$ induces backscattering. In this case, the KFP fixed point $\tilde g_{\tau \tau'}^i = 0$ is reached naturally as a consequence of disorder~\cite{KaneFisherPolchinski1994}.

\subsection{Interaction and hopping terms in the comoving frame}

As mentioned in Sec.~\ref{sec:DirtyReview}, the disorder potential can be absorbed by comoving fields under a unitary (viz. orthogonal) rotation defined by Eq.~\eqref{eq:hValuesForGeneralDisorder}.
The price to pay to gauge away the disorder Hamiltonian $H_\xi$ is that the disorder will reappear in the interedge couplings. Transforming the dipole interaction Eq.~\eqref{eq:Hsigma} yields
\begin{subequations}
\begin{equation}
    H_{\sigma}^{tb}=2\pi v_\sigma \int \dd x\,\mathbf J_t^{\mathrm T}\Lambda\mathbf J_b = 2\pi v_\sigma \int \dd x\,\widetilde{\mathbf J}_t^{\mathrm T}S_t^{\mathrm T}\Lambda S_b\widetilde{\mathbf J}_b,
\label{eq:disorderedDipolInteraction}
\end{equation}
where
$\mathbf J_\tau=\left(J_\tau^x,J_\tau^y,J_\tau^z\right)^{\mathrm T}$, and $\Lambda=\operatorname{diag}\left(\lambda_x,\lambda_y,\lambda_z\right)$. 
At the same time, it is a major feature of the comoving frame that the charge-dipole interaction Eq.~\eqref{eq:rhosigmaIA} looks effectively $x$-dependent in the comoving frame
\begin{align}
H_{\rho \sigma} &= \sum_{\substack{\tau, \tau' = t,b, \\ i=x,y,z}} \int \dd x\tilde{g}_{\tau \tau'}^i \partial_x \phi_{\rho,\tau} J_{\tau'}^i \notag\\
&=\sum_{\substack{\tau, \tau' = t,b, \\ i=x,y,z}} \int \dd x\tilde{g}_{\tau \tau'}^i \partial_x \phi_{\rho,\tau} [S_{\tau'} {\tilde{\mathbf J}}_{\tau'}]^i,
\end{align}
and eventually averages to zero~\cite{KaneFisherPolchinski1994}, $\overline{H_{\rho \sigma}} = 0$ in all disorder scenarios discussed below.

The tunneling and pairing Hamiltonian, respectively, transform as

\begin{multline}
    H_t = -\int \dd x\left(\Psi_t^\tau\mathcal T\Psi_b + \text{H.c.}\right) \\
    =-\int \dd x\left(\tilde\Psi_t^\dagger \hat U^\dagger_t\mathcal T \hat U_b \tilde \Psi_{b}+\text{H.c.}\right) \\
    = - \sum_{\alpha = 0}^3\int \dd x t_\alpha \left(\tilde\Psi_t^\dagger \hat U^\dagger_t\sigma_\alpha \hat U_b \tilde \Psi_{b}+\text{H.c.}\right),
\end{multline}
and
\begin{multline}
    H_\Delta = \int \dd x \left(\Psi^\text{T}_t \mathcal D \Psi_b + \text{H.c.}\right) \\ 
    =\int \dd x\left(\tilde\Psi_t^\text{T}\hat U^T_t\mathcal D \hat U_b \tilde\Psi_b + \text{H.c.}\right) \\
    = \sum_{\alpha = 0}^3\int \dd x \left(id_\alpha\sigma_y\tilde\Psi^\dagger_t \hat U^\dagger_t \sigma_\alpha \hat U_b\tilde\Psi_b + \text{H.c.}\right).
\end{multline}\label{eq:AlldisorderedDipole}
\end{subequations}

\subsection{Discussion of a variety of disorder scenarios}
\label{sec:DiscussionOfAVarietyOfDisorderScenarios}

We consider a total of three situations, each of which corresponds to a different physical realization.

On the one hand, we consider the situation in which the interaction potentials $\lambda_i(x), \tilde g_{\tau, \tau'}^i(x)$ as well as the hopping and CAR matrices $\mathcal T(x), \mathcal D(x)$ are themselves smooth functions and may be approximated by a constant. For this scenario of sharply defined trenches, we discuss both correlated and uncorrelated disorder potentials in Secs.~\ref{sec:CorrelatedDis}, ~\ref{sec:UNCorrelatedDis}.

On the other hand, we consider the situation in which hopping and CAR matrices $\mathcal T(x), \mathcal D(x)$ are rugged and partly compensate the random SU(2) matrices $\hat U_{t,b}$ appearing in ``comoving frame'' Eq.~\eqref{eq:AlldisorderedDipole}. Here, the distinction between correlated and uncorrelated disorder is less prominent, and the behavior is akin to the clean case, see Sec.~\ref{sec:RuggedHopping}. We discuss both the situation where $\lambda_i(x), \tilde g_{\tau, \tau'}^i(x)$ are smooth (they stem from Coulomb interactions and are less susceptible to random variations in trench thickness $d$) and where interactions are also fast random variables.

\subsection{Correlated disorder}
\label{sec:CorrelatedDis}

Studying the influence of generic disordered inter-edge couplings on the stability of parafermions is a complex problem. However, for the case of edge correlated disorder, we show that the same arguments as presented in Sec.~\ref{sec:cleanCase} hold, however without the assumption of fine tuning as the $\tilde g_{\tau, \tau'}^i$ terms vanish in the disorder average.

\subsubsection{Dipole interaction in the presence of correlated disorder}

We consider the physically relevant case $\xi_t=\xi_b^*$, which can arise when the random couplings on both edges originate from a common disorder potential, such as that produced by Coulomb impurities, dislocations, or strain. In this case, the coefficients $h_\tau^i$ in Eq.~\eqref{eq:hValuesForGeneralDisorder} simplify, and the gauge transformations on the two edges satisfy
\begin{equation}
S_t=\eta S_b\eta, \qquad \eta=\operatorname{diag}(-1,1,-1).
\end{equation}
Under disorder averaging, this relation projects the dipole interaction onto the direction $(-1,1,-1)$ (blue trajectory in Fig.~\ref{fig:FlowDiagrams}b), yielding
\begin{equation}
\overline{H_\sigma^{tb}} =2\pi v_\sigma\bar\lambda \int \dd x \widetilde{\mathbf J}_t^{\mathrm T} \eta \widetilde{\mathbf J}_b,
\label{eq:disorderAveragedDipol}
\end{equation}
where
$\bar\lambda=\frac{-\lambda_x+\lambda_y-\lambda_z}{3}. $
The averaged coupling obeys the RG equation
\begin{equation}
\frac{ d\bar\lambda}{d\ell}
=\bar\lambda^2+c\mathscr D_\sigma,
\label{eq:RGflowLambdaBar}
\end{equation}
where $c>0$ is a nonuniversal constant and $\mathscr D_\sigma$ characterizes the residual fluctuations around the averaged interaction. Their strength can be expected to scale as
\begin{equation}
\mathscr D_\sigma \sim \frac{v_\sigma^2}{W}\sum_i (
\eta_{ii}\lambda_i - \bar \lambda)^2,
\end{equation}
where $\mathscr{D}_{\sigma}$ is small/large when the projection of the vector $(\lambda_x, \lambda_y, \lambda_z)$ on $\bar\lambda (-1, 1,- 1)$ is small/large. 

Neglecting these fluctuations, the averaged interaction is marginally relevant for $\bar\lambda>0$ and marginally irrelevant for $\bar\lambda<0$. Thus, within this approximation, the half-space $-\lambda_x+\lambda_y-\lambda_z>0$ flows toward the strong-coupling ray $\lambda(-1,1,-1)$. Correlated disorder, therefore, changes the basin of attraction relative to the clean theory. In particular, an initial interaction $\boldsymbol{\lambda}=\lambda(-1,-1,-1)$, with $\lambda>0$, which flows to zero in the clean theory, has $\bar\lambda=\lambda/3>0$ and flows toward strong coupling in the disorder-averaged description.

Residual fluctuations modify this growth. Their contribution to Eq.~\eqref{eq:RGflowLambdaBar} is negligible when $\mathscr D_\sigma\ll\bar\lambda^2$, whereas comparable contributions can modify the basin of attraction depending on the sign of $c$. As for the charge–dipole couplings $\tilde g^i_{\tau\tau'}$, these fluctuations are RG irrelevant in the perturbative regime. Nevertheless, they can affect the basin of attraction through their feedback on $\bar\lambda$. Determining its boundary, therefore, requires the coupled flow of $\bar\lambda$ and $\mathscr D_\sigma$; the condition $(\bar\lambda>0)$ with $|c|\mathscr D_\sigma\ll\bar\lambda^2$ identifies the regime in which the fluctuation correction is initially small.

\subsubsection{Tunneling/CAR in the presence of correlated disorder}

In the fundamental representation, the top and bottom gauge transformations are related by
\begin{equation}
    \hat U_t = \sigma_x\hat  U_b\sigma _x.
\end{equation}
Thus, a crucial consequence of edge-correlated disorder is that the $x$-channel of the tunneling and pairing terms remains disorder invariant. For instance, for the tunneling we get 
\begin{equation}
    t_x(x) = t_x \hat U^\dagger_t  \sigma_x \hat{U}_b = t_x U_t^\dagger \sigma_x \sigma_xU_t\sigma_x = t_x\sigma_x.
\end{equation}
The same calculation holds for the CAR channel, too. Hence, it remains to determine how disorder affects the $0$, $y$, and $z$ CAR/tunneling channels.
We follow the Giamarchi-Schulz approach \cite{GiamarchiSchulz1988}, to derive an effective replica action
\begin{widetext}
\begin{equation}
\begin{aligned}
\, & \mathcal S_{\mathrm{rep}}
= 
\sum_n \mathcal S_\rho[\Theta_\rho^n, \Phi_\rho^n]+\sum_n \mathcal S_\sigma[\tilde\Theta^n_\sigma, \tilde\Phi^n_\sigma]\\ & 
-\frac{2t_x}{\pi a}
\sum_n
\int \ddd{x}{\tau}\,
\sin(\tilde\Phi_\sigma^n)
\cos(3\Theta_\rho^n)
 -
\frac{2|d_x|}{\pi a}
\sum_n
\int \ddd{x}{\tau}\,
\sin(\tilde\Phi_\sigma^n)
\cos(3\Phi_\rho^n+\delta_x)
\\
&-
\frac{\mathscr D_\Theta}{2\pi a^2}
\sum_{nm}
\int \mathrm{d}\tau\,\mathrm{d}\tau'\,\mathrm{d}x\,
\cos(\tilde\Theta_\sigma^n-\tilde\Theta_\sigma^{\prime m})
\cos(3\Theta_\rho^n-3\Theta_\rho^{\prime m})
-
\frac{\mathscr D_\Phi}{2\pi a^2}
\sum_{nm}
\int \mathrm{d}\tau\,\mathrm{d}\tau'\,\mathrm{d}x\,
\cos(\tilde\Phi_\sigma^n)
\cos(\tilde\Phi_\sigma^{\prime m})
\cos(3\Theta_\rho^a-3\Theta_\rho^{\prime m})
\\
&-
\frac{\mathscr C_\Theta}{2\pi a^2}
\sum_{nm}
\int \mathrm{d}\tau\,\mathrm{d}\tau'\,\mathrm{d}x\,
\cos(\tilde\Theta_\sigma^n-\tilde\Theta_\sigma^{\prime m})
\cos(3\Phi_\rho^n-3\Phi_\rho^{\prime m})
-
\frac{\mathscr C_\Phi}{2\pi a^2}
\sum_{nm}
\int \mathrm{d}\tau\,\mathrm{d}\tau'\,\mathrm{d}x\,
\cos(\tilde\Phi_\sigma^n)
\cos(\tilde\Phi_\sigma^{\prime m})
\cos(3\Phi_\rho^n-3\Phi_\rho^{\prime m}), 
\end{aligned}
\label{eq:ReplicaAction}
\end{equation}
\end{widetext}
where $\mathcal S_\rho$ and $\mathcal{S}_\sigma$ are the actions that correspond to the Hamiltonians $H_\rho$ and the disorder averaged $\overline{H_\sigma}$. Here, $n,m$ are replica indices.

The effective disorder strengths $\mathscr D_{\Phi,\Theta}$ and $\mathscr C_{\Phi,\Theta}$ have same dimensions as $W$ and scale as
\begin{equation}
    \mathscr D_{\Phi,\Theta}
    \sim
    \frac{v_\sigma^2}{W}\bar t^2,
    \qquad
    \mathscr C_{\Phi,\Theta}
    \sim
    \frac{v_\sigma^2}{W}\bar d^2,
\end{equation}
where $\bar t$ and $\bar d$ are weighted mean-square averages of the couplings $t_{0,y,z}$ and $d_0,d_{y,z}$, respectively. Further details are given in App.~\ref{secApp:replicaAction}. The tilde on the $\sigma$ fields indicates that they are written in the transformed frame in which the original disorder has been absorbed into the fields.

In contrast to the clean case, the effective replica action \eqref{eq:ReplicaAction} contains only a single hopping and a single pairing term. At the same time, it includes the four additional $\mathscr D_{\Theta,\Phi}$ and $\mathscr C_{\Theta,\Phi}$ couplings describing the mean-square of backscattering and pairing, respectively, coupling different replica fields. Thus, to infer the low-energy physics of the whole system, the tunneling, the pairing, and the exchange coupling should also be compared with the effective disorder terms. The RG flow equations for $\lambda_{x,y,z}$ in Eq.~\eqref{eq:RGFlowDipol}, for $\mathscr t_x$ in Eq.~\eqref{eq:RGFlowTx}, and the analogous equation for $\mathscr d_x$, must be complemented by the corresponding flow equations (in the weak exchange regime)
\begin{subequations}
\begin{align}
    \frac{d\mathscr D_\Theta}{d\ell}
    &=
    \left(
    2-3K_\rho+\frac{\lambda_z}{2}
    \right)
    \mathscr D_\Theta,
    \\
    \frac{d\mathscr D_\Phi}{d\ell}
    &=
    \left(
    2-3K_\rho-\frac{\lambda_z}{2}
    \right)
    \mathscr D_\Phi,
    \\
    \frac{d\mathscr C_\Theta}{d\ell}
    &=
    \left(
    2-\frac{3}{K_\rho}+\frac{\lambda_z}{2}
    \right)
    \mathscr C_\Theta,
    \\
    \frac{d\mathscr C_\Phi}{d\ell}
    &=
    \left(
    2-\frac{3}{K_\rho}-\frac{\lambda_z}{2}
    \right)
    \mathscr C_\Phi.
\end{align}
\end{subequations}

At the special case of no inter-edge interactions, i.e., $K_\rho=1$ and $\lambda_z=0$, all disorder terms are strongly irrelevant. Along the strong-coupling ray $\boldsymbol{\lambda} \rightarrow \lambda(-1, 1,-1)$ that leaves the $x$ channels invariant, the disorder couplings $\mathscr D_\Phi$ and $\mathscr C_\Phi$ are favored since $\lambda_z <0$. The green hatched region in Fig.~\ref{fig:PhaseDiagram}b) shows where both of these couplings remain irrelevant.

Interestingly, most of the parafermion phase lies within this region. Thus, throughout a large part of the parameter regime in which parafermions are expected, disorder does not affect the low-energy physics. Outside this region, at least one of the two remaining disorder couplings becomes RG relevant. Even then, however, as long as the system lies above the black line, the exchange coupling reaches strong coupling first and pins the neutral field. The relevant disorder contribution does not survive once the neutral sector is integrated out. Hence, it does not introduce any further complication into the effective low-energy theory.

At this point, it is important to stress that the entire mechanism relies on the crucial assumption of edge-correlated disorder. This particular correlation enables a strong-coupling ray and leaves a tunneling/pairing channel invariant. Moreover, disorder-invariant CAR/tunneling channel is exactly the one favored by the strong coupling ray. This results again in an effective low-energy double sine-Gordon theory and consequently in PZMs.

\subsection{Uncorrelated disorder}
\label{sec:UNCorrelatedDis}

The situation changes when the disorder on the top and bottom edges is completely uncorrelated, $\overline{\xi_t(x)\xi_b^*(x')}=0.$
In this case, any inter-edge coupling averages to zero at first order. This applies not only to the dipole interaction \eqref{eq:disorderedDipolInteraction} but also to the pairing and hopping terms. In particular, to leading order in $t_\alpha$, $d_\alpha$, and $\lambda_{x,y}$, only inter-replica interactions can be generated.

If relevant, such interactions need not drive the system into a fully gapped phase. We expect that one possibility is an Anderson-localized, Bose-glass-like state with a mobility gap rather than a spectral gap \cite{FisherFisher1989}.  
The fields $\Phi_\rho$ and $\Theta_\rho$ could remain locally pinned within their respective regions, with pinning phases $\alpha(x)$ that vary with the disorder potential, rather than taking the uniform values associated with correlated disorder. Whether such pinning can support parafermionic bound states, and how residual low-energy excitations would affect their localization and observable signatures, remain open questions requiring further investigation.

Still, the whole construction relies on the assumption that both pairing and tunneling are relevant within the same region of parameter space $(K_\rho,\lambda_z^0)$. In the scenario considered here, this is not possible. For example, suppose that the low-energy theory is dominated by $\mathscr D_\Phi$ and $\mathscr C_\Phi$, while the dipole interaction can be neglected. The corresponding effective replica action is then analogous to the action in Eq.~\eqref{eq:ReplicaAction}, except that the terms linear in $t_\alpha$ and $d_\alpha$ are absent, $\mathcal S_\sigma$ contains only the free quadratic action of the neutral modes, and $\bar t$ and $\bar d$ collect all couplings $t_\alpha$ and $d_\alpha$, respectively.

As shown in Fig.~\ref{fig:PhaseDiagram}b), the regions in which $\mathscr D_\Phi$ and $\mathscr C_\Phi$ are relevant are separated by the green hatched region, in which neither coupling is relevant. Therefore, there is no region of parameter space in which both $\Theta_\rho$ and $\Phi_\rho$ can be pinned simultaneously.

Let us further assume that the neutral field is pinned by some additional, unspecified mechanism, for instance by the replica-generated counterpart of the dipole interaction. In that case, one expects an effective replica action of the form
\begin{equation}
\begin{split}
\mathcal S^\text{eff}_\text{rep}
= & \sum_a \mathcal S_\rho[\Phi_\rho^a,\Theta_\rho^a] \\
& - \frac{\mathscr D_\text{eff}}{2\pi a^2}
\sum_{ab}\int  \text{d}\tau\text{d}\tau'\text{d}x\,
\cos\left(3\Theta_\rho^a-3\Theta'^b_\rho\right)\\
& - \frac{\mathscr C_\text{eff}}{2\pi a^2}
\sum_{ab}\int \text{d}\tau\text{d}\tau'\text{d}x\,
\cos\left(3\Phi_\rho^a-3\Phi'^b_\rho\right),
\end{split}
\end{equation}
with the corresponding flow equations
\begin{subequations}
\begin{align}
\frac{d\mathscr D_\text{eff}}{d\ell}
&=
\left(3-3K_\rho\right)\mathscr D_\text{eff},\\
\frac{d\mathscr C_\text{eff}}{d\ell}
&=
\left(3-\frac{3}{K_\rho}\right)\mathscr C_\text{eff}.
\end{align}
\end{subequations}
The two couplings, therefore, cannot be simultaneously relevant: $\mathscr D_\text{eff}$ is relevant for $K_\rho<1$, whereas $\mathscr C_\text{eff}$ is relevant for $K_\rho>1$. At the noninteracting point $K_\rho=1$, both couplings are marginal.

We therefore conclude that, for completely uncorrelated disorder on the two edges, the simultaneous pinning of $\Theta_\rho$ and $\Phi_\rho$ required for the parafermionic construction cannot be achieved. Consequently, we do not expect stable PZMs in this regime.

\subsection{Rugged hopping and CAR}
\label{sec:RuggedHopping}

We now consider the case in which disorder also affects the microscopic coupling amplitudes. As discussed in Sec.~\ref{sec:DiscussionOfAVarietyOfDisorderScenarios}, we assume that the tunneling and CAR amplitudes, $\mathcal T(x)$ and $\mathcal D(x)$, contain spatial fluctuations that are partially compensated by the local SU(2) rotations $\hat U_{t,b}(x)$. In the rotated basis, these amplitudes take the form
\begin{equation}
    \hat U_t^\dagger \mathcal T(x) \hat U_b
    = \overline{\mathcal T} + \delta\mathcal T(x),
    \quad
    \hat U_t^\dagger \mathcal D(x) \hat U_b
    = \overline{\mathcal D} + \delta\mathcal D(x),
\end{equation}
where $\overline{\mathcal T}$ and $\overline{\mathcal D}$ are constant matrices, while $\delta\mathcal T(x)$ and $\delta\mathcal D(x)$ are zero-mean, short-range-correlated fluctuations. Since the random contributions are less relevant under the RG flow than the uniform terms, we retain only the latter. The tunneling and CAR Hamiltonians therefore reduce, at leading order, to their clean-system forms in Eq.~\eqref{eq:tDeltaBosonized}. Note that $\overline{H_{\rho \sigma}} = 0$ holds for the rugged trench, as well, so that no fine tuning of $\tilde g_{\tau \tau'}^i$ is needed.

First, we consider the case in which the interaction couplings $\lambda_i(x)$, however, are still assumed to vary smoothly in space. Consequently, after absorbing the disorder into the current operators, the dipole interaction $H_\sigma^{tb}$ must still be disorder-averaged. For uncorrelated disorder on the two edges, this average vanishes to first order, leaving no channel-selection mechanism at this order. For correlated disorder, by contrast, the averaged dipole interaction takes the form of Eq.~\eqref{eq:disorderAveragedDipol}. When the surviving coupling $\bar\lambda$ drives the neutral sector toward the strong-coupling ray $\lambda(-1,1,-1)$, the interaction selects the tunneling and CAR $x$-channel, leading to PZMs. In the limit $\bar\lambda\to0$, the RG relevance of tunneling and CAR is instead controlled solely by $K_\rho$, and the two processes cannot be simultaneously relevant.

Finally, if the dipole interaction also contains microscopic disorder that is compensated by the same local rotations, its leading contribution likewise recovers the clean-system form. The full model then reduces, at leading order, to the clean case discussed in Sec.~\ref{sec:cleanCase}.

\section{Conclusion}
\label{sec:Conclusion}

We have developed a theory delineating regimes for the realization of $\mathbb{Z}_3$ parafermionic zero modes in spin-polarized $\nu=2/3$ fractional quantum Hall heterostructures. Unlike in simpler spin-unpolarized $\nu=2/3$ Halperin or in Laughlin state constructions, tunneling and superconducting pairing 
\textit{a priori} generate several competing operators involving both charge and neutral degrees of freedom. The central result of this work is that inter-edge interactions can dynamically resolve this competition and reduce the low-energy theory to the double sine-Gordon model \eqref{eq:reviewBosonizedTunnelingPairing} 
hosting PZMs at domain walls.

The mechanism is summarized in Fig.~\ref{fig:Mechanisim}. As the system approaches the KFP fixed point, the microscopic electron and $e/3$ quasihole modes reorganize into a charged mode carrying charge $-2e/3$ and a neutral mode with an internal dipole degree of freedom. The emergent ${\mathfrak{su}}(2)_1$ structure of the neutral sector allows to identify inter-edge dipole-exchange interactions that can become important under the RG flow. When these interactions reach strong coupling first, they pin the neutral field and gap out the neutral dipole sector 
and leave the effective charge-sector double sine-Gordon Hamiltonian \eqref{eq:effectiveDoubleSineGordon} behind.

We first established this mechanism in a clean model, where the separation of charged and neutral modes requires (at least approximate) fine-tuning of the inter- and intra-edge interactions. We investigated the role of dipole exchange interactions and showed that their presence substantially enlarges the phase hosting PZMs, Fig.~\ref{fig:PhaseDiagram} a), b).  

We further studied the disordered case where intra-edge backscattering provides a more generic route to the same KFP regime, where no fine-tuning is needed. 
We considered different disorder scenarios. In the first scenario, we assumed that the microscopic tunneling, CAR, and interaction strengths are approximately constant and become random variables when transforming into the frame comoving to the disorder.
Most importantly, in the case of edge-correlated disorder, we identified that a specific strong-coupling fixed point of the neutral modes as well as one tunneling and one pairing channel, remains unaffected by disorder, allowing the neutral sector to be gapped without destroying the effective parafermion generating channel. We demonstrate that the residual disorder is itself RG irrelevant, ensuring a stable PZM phase in a substantial part of the resulting phase diagram, Fig.~\ref{fig:PhaseDiagram} c). 
Last, we discussed the effect of a rugged interface with microscopically random tunnel and CAR couplings that, themselves, become constant in the comoving frame, and the system behaves similarly to the clean case. 

These results show that the multicomponent structure of the polarized $\nu=2/3$ edge need not prevent parafermion formation. Instead, the neutral dipole sector provides the mechanism that selects the appropriate low-energy tunneling and pairing processes. We identified the important parameter spanning a broad parameter space in which the effective fractional charge theory supports $\mathbb{Z}_3$ PZMs, without requiring direct fine-tuning of the microscopic tunneling and superconducting channels. As such, we are optimistic that gate-defined, single material heterostructures of FCI-SC interfaces may serve as realistic platforms towards the realization of $\mathbb Z_3$ parafermionic zero modes.

\section*{Note added}

During the final stages of preparing this manuscript, an independent study \cite{Amire2026} on the same topic was posted on the arXiv.

\section*{AI disclosure}

In the later stages of the project, the authors used ChatGPT 5.6 and ChatGPT 6.1 (OpenAI) for literature search and checking calculations,
and take full responsibility for the content.

\acknowledgments

It is a pleasure to acknowledge useful discussions with J. Cao, E. Fradkin, A.D. Mirlin, Y. Oreg, P. M. Bonetti, and D. Vilardi. 
J.I.V.~was supported by the National Science Foundation award number DMR-2540313. T.L.S.~and A.H.~acknowledge financial support from the National Research Fund of Luxembourg under Grants C22/MS/17415246/DeQuSky and AFR/23/17951349. Support for this research was provided by the
Office of the Vice Chancellor for Research and Graduate Education at the University of Wisconsin–Madison with funding
from the Wisconsin Alumni Research Foundation (E.J.K.). This work was performed in part at the Aspen Center for Physics, which is supported by National Science Foundation grant PHY-2210452 (J.I.V. and E.J.K.).

\appendix

\section{(Non-Abelian) Bosonization of the Tunneling and Pairing Term}
\label{secApp:Bosonization}

In this Appendix we provide details behind tunneling and pairing terms Eq.~\eqref{eq:BosonizedHt},\eqref{eq:BosonizedHDelta}. We use the two most and equally RG relevant electron operators at the KFP fixed point, Eq.~\eqref{eq:PMfermionDefinition} and their spinor representation Eq.~\eqref{eq:PMfermionDefinition}. Moreover, we remind the reader that the most relevant electron operator at the KFP fixed point is a superposition, Eq.~\eqref{eq:electronOperator}.

\subsection{Tunneling}

The generic tunneling term reads, cf.~Eqs.~\eqref{eq:electronOperator},\eqref{eq:SpinorHopt} 
\begin{multline}
     H_t = -t \int \dd x \left(\psi_t^\dagger \psi_b + \text{H.c.}\right) \\= -t\int \dd x \left(\Psi_t^\dagger \mathcal T\Psi_b + \text{H.c.}\right). 
\end{multline}
In the following, we make use of that
\begin{equation}
    \Psi_t^\dagger \sigma_\alpha \Psi_b = - \text{tr}\left[\sigma_\alpha (\Psi_b\otimes\Psi_t^\dagger)\right], 
\end{equation}
where the symbol
\begin{equation}
    (\Psi_b\otimes \Psi^\dagger_t) = \begin{pmatrix}
        \psi_{b+}\psi^\dagger_{t+} & \psi_{b+}\psi^\dagger_{t-} \\
        \psi_{b-}\psi^\dagger_{t+} & \psi_{b-}\psi^\dagger_{t-}
    \end{pmatrix}
\end{equation}
denotes the tensor product between the spinor on the top and bottom edge. This matrix can further be expanded in terms of the spinor fields, Eq.~\eqref{eq:SpinorRep}, $\v z_\tau$ via
\begin{equation}
    (\Psi_b\otimes \Psi^\dagger_t) = \frac{i\sqrt{2}}{2\pi a} e^{-i3\Theta_\rho}\underbrace{D_\kappa : \frac{(\v z_b\otimes \mathbf z_t^\dagger)}{\sqrt{2}}: D_\kappa.}_{g}
\end{equation}
where
\begin{equation}
    g = \begin{pmatrix}
        e^{-i\Theta_\sigma} & e^{i\Phi_\sigma} \\
        -e^{-i\Phi_\sigma} & e^{i\Theta_\sigma}
    \end{pmatrix}
\end{equation}
is an operator valued SU(2) matrix. 
The matrix $\frac{1}{\sqrt 2}\mathbf z_b\otimes \v z_t^\dagger\equiv \mathscr g$ transforms under a pseudospin rotation as 
\begin{equation}
    \mathscr g = U_b \tilde{\mathscr{g}}U_t^\dagger.  
\end{equation}
Consequently, the matrix $g$ transforms accordingly 
\begin{equation}
   g = \hat U_b \,\tilde g\, \hat U_t^\dagger,\text{ where } \hat{U}_\tau = D_\kappa U_\tau D_\kappa.
\end{equation}
Inserting the $g$ representation of the fermion bilinear yields 
\begin{equation}
    -\text{tr}\left[\sigma_i (\Psi_b\otimes \Psi_t^\dagger)\right] = \frac{-i\sqrt{2}}{2\pi a}\text{tr}[\sigma_i g] e^{-i3\Theta_\rho}. 
\end{equation}
Thus, we simply need to calculate the trace of $g$ to have the bosonized Hamiltonian:
\begin{subequations}
    \begin{align}
        \text{tr}[\sigma_0 g] & = \sqrt2\cos(\Theta_\sigma), \\
        \text{tr}[\sigma_x g] & = \sqrt2 i\sin(\Phi_\sigma), \\
        \text{tr}[\sigma_y g] & = \sqrt 2 i \cos(\Phi_\sigma), \\
        \text{tr}[\sigma_z g] & = -\sqrt 2 i \sin(\Theta_\sigma). 
    \end{align}
    \label{AppEq:tracesOverg}
\end{subequations}
We note that $i\text{tr}[\sigma_i g]\in \mathbb R$ for $i=x,y,z$. Thus, we can compactly write the tunneling Hamiltonian as 
\begin{multline}
     H_t = \frac{\sqrt 2t_0}{\pi a} \int \dd x \,\text{tr}[g]\cos(3\Theta_\rho) \\ + \frac{\sqrt{2}}{\pi a}\v t \cdot \int \dd x \,i\text{tr}[\boldsymbol\sigma g]\sin(3\Theta_\rho),
\end{multline}
where $\v t = (t_x,t_y,t_z)$. In the fully bosonized form, the Hamiltonian reads
    \begin{align}
         H_t &= \frac{2t_0}{\pi a}\int \dd x \cos(\Theta_\sigma)\sin(3\Phi_\rho) \notag\\ & + \frac{2}{\pi a}\int \dd x \left(t_z \sin(\Theta_\sigma)-t_x\sin(\Phi_\sigma)-t_y\cos(\Phi_\sigma)\right)\notag\\ &\times \cos(3\Theta_\rho).
    \end{align}
This concludes the derivation of Eq.~\eqref{eq:BosonizedHt}.

\subsection{Pairing}
The same analysis can be carried out for the pairing part of the Hamiltonian. For that, we define the superconducting order parameter as $ \mathcal D = (d_0 +  \v d\cdot \boldsymbol\sigma)i\sigma_y$, where the values of $d_0$ and $\mathbf d = (d_x,d_y,d_z)$ are determined by the coefficients $c_{\tau,\pm}$. Thus, we can expand the pairing term as
\begin{multline}
     H_\Delta = \Delta \int \dd x \left(\psi_t\psi_b +\text{H.c.}\right)\\ = -i\int \dd x \big(d _0 \text{tr}[\sigma_y (\Psi_b\otimes \Psi_t)] \\+ \v d\cdot \text{tr}[\boldsymbol\sigma\sigma_y(\Psi_b \otimes\Psi_t)] + \text{H.c.}\big).
\end{multline}
It is straightforward to show that 
\begin{equation}
    \sigma_y(\Psi_b\otimes \Psi_t^T) = -\frac{\sqrt{2}}{2\pi a}e^{i3\Phi_\rho}g^T.
\end{equation}
Again, the full pairing part is fully bosonized once the traces $\text{tr}[\sigma_i g^T]$ are known. Since the transpose only acts as a sign change in the $\sigma_y$ component of $g$, we use the results as in \eqref{AppEq:tracesOverg} but changing the sign of the $y$ component. Thus the fully bosonized pairing part becomes 
\begin{widetext}
    \begin{equation}
    \begin{split}
        H_\Delta  = & -\frac{2d_0}{\pi a}\int \dd x \cos(\Theta_\sigma)\sin(3\Phi_\rho) - \frac{2|d_x|}{\pi a} \int \dd x \sin(\Phi_\sigma)\cos(3\Phi_\rho + \gamma_x) \\
        & + \frac{2|d_y|}{\pi a}\int \dd x\cos(\Phi_\sigma)\cos(3\Phi_\rho + \gamma_y) + \frac{2|d_z|}{\pi a}\int \dd x \sin(\Theta_\sigma)\cos(3\Phi_\rho+\gamma_z)
    \end{split}
    \end{equation},
\end{widetext}
where $d_i = |d_i|e^{i\gamma_i}$, see Eq.~\eqref{eq:BosonizedHDelta}. Note that we chose the phase gauge of the order parameter $\mathcal D$ such that $d_0 \in \mathbb R$. From here, it is clear that the $H_\Delta$ can also be easily expressed in terms of traces over $g$ instead of $g^T$. When absorbing a minus sign into the phase $\gamma_y \rightarrow\gamma_y + \pi$, the pairing term is conveniently expressed as
\begin{multline}
     H_\Delta = - \frac{\sqrt2 d_0}{\pi a} \int \dd x \,\text{tr}[g]\sin(3\Phi_\rho)\\ + \frac{\sqrt 2}{\pi a} \sum_{j=x,y,z} |d_j|\int \dd x \, i\,\text{tr}[\sigma_j g]\cos(3\Phi_\rho +\gamma_j).
\end{multline}

\section{Free neutral fields: Parafermions without Exchange Interactions}
\label{secApp:parafermionsVanila}

When the dipole interactions approach a Gaussian fixed point, i.e. $(\lambda_x, \lambda_y,\lambda_z)\rightarrow(0,0,\pm \lambda)$ or an equivalent fixed point, formally all terms of Eq.~\eqref{eq:BosonizedHt},\eqref{eq:BosonizedHDelta} contribute. Yet, the sign of $\lambda_z$ will either propel the $\Theta_\sigma$ or the $\Phi_\sigma$ channel under RG. This leads to a reduction of terms and ultimately to the stabilization of parafermionic modes and is the topic of this appendix.

We concentrate on $\lambda_z >0$. In this case, keeping only the RG-favored terms, the tunneling part Eq.~\eqref{eq:BosonizedHt} simplifies to 
\begin{align}
    \,  & H_t 
    = \frac{2}{\pi a}\int\dd x\left(t_+ \sin(\chi_+)+t_-\sin(\chi_-)\right), 
\end{align}
where 
\begin{align}
    \chi_+ & = 3\Theta_\rho + \Theta_\sigma,\quad t_+ = \frac{t_0 + t_z}{2} \\
    \chi_- & = 3\Theta_\rho - \Theta_\sigma,\quad t_- = \frac{t_0 - t_z}{2}.
\end{align}
Semiclassically, the fields are pinned to 
\begin{align}
        \braket{\chi_+} & = 3\braket{\Theta_\rho} + \braket{\Theta_\sigma} = \underbrace{-\frac{\pi}{2}\text{sign}(t_+)}_{\delta_+} + 2\pi\hat n_+, \\
        \braket{\chi_-} & = 3\braket{\Theta_\rho} - \braket{\Theta_\sigma} = \underbrace{-\frac{\pi}{2}\text{sign}(t_-)}_{\delta_-} + 2\pi\hat n_-,
\end{align}
where $\hat n_\pm$ are integer-valued operators.
Solving for the expectation values of the original fields yields
\begin{align}
    \braket{\Theta_\rho} & = \frac{\pi}{3}\underbrace{(\hat n_+ + \hat n_-)}_{\hat{\tilde m}} + \frac{1}{6}(\delta_++\delta_-) \\
    \braket{\Theta_\sigma} & = \pi\underbrace{(\hat n_+ - \hat n_-)}_{\hat M} + \frac{1}{2}(\delta_+-\delta_-).
\end{align}
Note that the new integer values operators $\hat{\tilde m}$ and $\hat M$ are not independent, i.e. $\hat M \text{ mod 2} = \hat{\tilde{m}}$. Without loss of generality, let us assume that $\delta_+ - \delta_- =0$ from which directly follows that $\frac{1}{6}(\delta_+ + \delta_-) = \mp\frac{\pi}{6}$, where the sign depends of the sign of $t_{0,z}$. In that case $\braket{\Theta_\sigma}=\pi \hat M$. Since both fields commute, we can fix $\Theta_\sigma$ first and chose $\Theta_\rho$ accordingly. Then we can shift $\hat{\tilde m}=2\hat m + \hat M$ so that the constraint on $\hat M$ and $\hat{\tilde m}$ is always fulfilled. Eventually, we find
\begin{subequations}
\begin{align}
    \Theta_\rho & = \frac{2\pi}{3}\hat m + \frac{\pi}{3}\hat M \pm \frac{\pi}{6}, \\
    \Theta_\sigma & = \pi\hat{ M}. 
\end{align}
\label{eqApp:pinnedThetaField}
\end{subequations}
Now we focus on the pairing Hamiltonian, which becomes for $\lambda_z>0$
\begin{align}
    H_\Delta 
    & =\frac{2}{\pi a}\int \dd x |Z(\Theta_\sigma)| \sin(3\Phi_\rho + \phi(\Theta_\sigma)),
\end{align}
where $Z(\Theta_\sigma)=i|d_z|\sin(\Theta_\sigma)e^{i\delta_z}-d_0 \cos(\Theta_\sigma)$ and $\phi(\Theta_\sigma)=\arg(Z(\Theta_\sigma))$. To minimize the energy, we need to maximize $|Z(\Theta_\sigma)|$ and pin the Sine to $-1$. For $\Theta_\sigma$, this yields
\begin{equation}
    \braket{\Theta_\sigma} = \pi \hat M + \underbrace{\frac{1}{2}\arctan\left(\frac{2 d_0 |d_z|\sin(\delta)}{d_0^2 - |d_z|^2}\right)}_{\varphi},
\end{equation}
In the next step, we pin the charge field to
\begin{equation}
    \braket{\Phi_\rho} = \frac{2\pi}{3}\hat n - \frac{1} {3}\phi(\braket{\Theta_\sigma})-\frac{\pi}{6}.
\end{equation}
We shift the neutral field exactly by the phase $\varphi$ and define $ \Theta \rightarrow \Theta - \varphi$, so that the neutral field is pinned to $\braket{\tilde\Theta_\sigma}=\pi \hat M$. From here, we note that
\begin{equation}
    Z(\braket{\Theta_\sigma})=Z(\pi \hat M + \varphi) = (-1)^{\hat M}Z(\varphi) = e^{i\pi \hat M}Z(\varphi).
\end{equation}
In particular, from here it follows that $\phi(\braket{\Theta_\sigma}) = \pi \hat M+\phi(\varphi)$. In the second step, we pin the charged field to minimize the sine function, and we obtain 
\begin{equation}
\begin{split}
    3\braket{\Phi_\rho}&  = 2\pi\hat n - \frac{\pi}{2} - \phi(\braket{\Theta_\sigma}) \\ & = 2\pi\hat n - \frac{\pi}{2} -\pi \hat M -\phi(\varphi).
\end{split}
\end{equation}
Eventually, we absorb the remaining phase $\phi(\varphi)$ in the the charge field $\Phi_\rho$ and we obtain the equations
\begin{subequations}
    \begin{align}
        \braket{\Phi_\rho} & = \frac{2\pi}{3}\hat n - \frac{\pi}{3}\hat M -\frac{\pi}{6}, \\
        \braket{\Theta_\sigma} & =\pi \hat M.
    \end{align}
\label{eqApp:pinnedPhiField}
\end{subequations}
Equations \eqref{eqApp:pinnedThetaField} and \eqref{eqApp:pinnedPhiField} are essentially the same equations (up to constant phase shifts) as in Ref.\cite{Mong2014}, in which the authors demonstrate that this structure, combined with the commutation relations of $\Theta_\rho$ and $\Phi_\rho$, yields $\mathbb Z_3$ PZMs. 

For $\lambda_z<0$, the $\Phi_\sigma$ sector will dominate, i.e. the $x$ and $y$ tunneling and pairing channels. It is straightforward to confirm that similar arguments hold.

\section{Pinned neutral field: Integration of massive fluctuations}

Contrary to the previous appendix, in this appendix, we concentrate on the case when the neutral field is pinned and discuss the formal integration thereof. For the example used in the main text, the strong coupling ray is  $(\lambda_x, \lambda_y, \lambda_z)\rightarrow\lambda(1, 1, 1)$, and we have 
\begin{equation}
\begin{split}
    S_\sigma  = & \frac{K_\sigma}{4\pi}\int \ddd{x}{\tau} \left[u_\sigma (\partial_x\Phi_\sigma)^2 + \frac{1}{u_\sigma}(\partial_\tau \Phi_\sigma)^2\right] \\
     & -  \frac{\lambda_b}{2\pi a^2}\int \ddd{x}{\tau} \cos(2\Theta_\sigma).
\end{split}
\end{equation}
The cosine pins the field, e.g. for $|\lambda_z^{(0)}|\gg|\lambda_p^{(0)}|$, with $\lambda_p$ defined in Eq.~\eqref{eq:lambdabp}, the bosons develop a gap $m_\sigma = T_{KFP}\left(2\frac{|\lambda_z^{(0)}|}{|\lambda_p^{(0)}|}\right)^{-1/|\lambda_p^{(0)}|}$. For $|\lambda_z^{(0)}|\sim|\lambda_p^{(0)}|$ the gap becomes
\begin{equation}
    m_\sigma = T_{\text{KFP}}\exp\left(-\frac{c}{|\lambda_b^{(0)}|}\right)
\end{equation}
with     \begin{equation}
        c = \int_r^\infty dx \frac{1}{x^2+1-r^2}
    \end{equation}
and $r=\frac{|\lambda_z^{(0)}|}{|\lambda_b^{(0)}|}$.
The field acquires a finite expectation, which is
\begin{equation}
    \braket{\Phi_\sigma} = \begin{cases}
        \pi n\quad n\in\mathbb Z, & \lambda_b >0, \\
        \pi n + \frac{\pi}{2}\quad n\in \mathbb Z, &\lambda_b <0.
    \end{cases}
\end{equation}
We define $\vartheta = \Theta_\sigma - \braket{\Theta_\sigma}$ as the small fluctuation around the expectation value and approximate 
\begin{equation}
    S_\sigma = \frac{1}{4\pi K_\sigma}\int \ddd{x}{\tau} \left[u_\sigma(\partial_x\vartheta)^2 + \frac{1}{u_\sigma}(\partial_\tau\vartheta)^2 +\frac{m_\sigma^2}{u_\sigma}\vartheta^2\right],
    \label{eqApp:gappedFreeNeutralAction}
\end{equation}
where $K_\sigma$ is not the bare value anymore but evaluated at the RG time $\ell^*$ where the neutral channel reaches strong coupling.

Now we are ready to integrate out the neutral sector in the tunneling and pairing terms
\begin{equation}
\begin{split}
    S_t + S_\Delta  = & + \frac{2}{\pi a}\int \dd xt_x\cos(\Theta_\sigma)\sin(3\Theta_\rho) \\ & + \frac{2}{\pi a}\int \dd xt_y\sin(\Theta_\sigma)\cos(3\Theta_\rho) \\
  & -\frac{2}{\pi a} \int \dd x |d_0|\cos(\Theta_\sigma)\sin(3\Phi_\rho) \\
  & +\frac{2}{\pi a} \int \dd x |d_z|\sin(\Theta_\sigma)\cos(3\Phi_\rho+\delta_z).
\end{split}
\end{equation}
Note that, since we assumed $\lambda_z >0$, we only kept the terms that are determining the low-energy physics in this case. The effective action is defined as
\begin{equation}
\begin{split}
    \, & S_\text{eff} = -\ln \left(\int \mathcal D \Phi_\sigma e^{-S_\sigma-S_t-S_\Delta}\right) \\
    & = \braket{S_t}_\sigma + \braket{S_\Delta}_\sigma - \frac{1}{2}\left(\braket{(S_t +S_\Delta)^2}_\sigma-\braket{S_t +S_\Delta}_\sigma^2\right)+\dots
    \label{eqApp:effectiveActionAfterNeutralMode}
\end{split}
\end{equation}
Essentially, we only need expectation values of the form $\braket{\cos(\Theta_\sigma)},\braket{\sin(\Theta_\sigma)}$. For that, we replace the field by its expectation value and the small fluctuations, i.e. for $\lambda_b > 0$
\begin{equation}
\begin{split}
   \braket{ \cos(\Theta_\sigma)}_\sigma & = \braket{\cos(\pi n +\vartheta)}_\sigma = e^{-\frac{1}{2}\braket{\vartheta^2}_\sigma}\cos(\pi n) \\ & = (-1)^n e^{-\frac{1}{2}\braket{\vartheta^2}_\sigma} = e^{i\braket{\Phi_\sigma}-\frac{1}{2}\braket{\vartheta}_\sigma},
\end{split}
\end{equation}
and 
\begin{equation}
    \braket{ \sin(\Theta_\sigma)}_\sigma  = \braket{\sin(\pi n +\vartheta)}_\sigma = e^{-\frac{1}{2}\braket{\vartheta^2}_\sigma}\sin(\pi n) = 0.
\end{equation}
Repeating the calculation for $\lambda_b <0$ reveals that the roles of sine and cosine are just reversed. Thus, the sign of $\lambda_b$ eventually determines which of the two tunneling/pairing channels survives integrating out the neutral mode. With action \eqref{eqApp:gappedFreeNeutralAction}, the equal space/time correlation function is
\begin{multline}
    \braket{\vartheta^2}_\sigma = -\frac{1}{2K_\sigma}\ln\left(\frac{M^2}{\Lambda^2 + M^2}\right) \\ \approx \frac{1}{K_\sigma}\ln\left(\frac{u_\sigma}{v_\sigma}\frac{T_\text{KFP}}{m_\sigma}\right),
\end{multline}
for $M\ll \Lambda$, where $M=m_\sigma/u_\sigma$. Thus, the correction factor to the coupling becomes
\begin{multline}
    e^{-\frac{1}{2}\braket{\vartheta^2}_\sigma} = \left(\frac{v_\sigma}{u_\sigma} \frac{m_\sigma}{T_\text{KFP}}\right)^{\frac{1}{2K_\sigma}} \\ = \left(\frac{v_\sigma}{u_\sigma}\right)^{\frac{1}{2K_\sigma}}\left(2\frac{|\lambda_z^0|}{|\lambda_b^0|}\right)^{-1/|\lambda_b^z|+\frac{1}{2K_\sigma}}.
\end{multline} in the small exchange limit and 
\begin{equation}
    e^{-\frac{1}{2}\braket{\vartheta^2}_\sigma} = \left(\frac{v_\sigma}{u_\sigma}e^{-c/|\lambda_b^0|}\right)^{\frac{1}{2K_\sigma}}
\end{equation}
in the strong exchange case.

Integrating out the neutral modes always retains the first-order tunneling and pairing terms. We therefore truncate equation \eqref{eqApp:effectiveActionAfterNeutralMode} at the first non-trivial order since $t_\alpha^2$ and $d_\alpha^2$ are parametrically small. 

The calculation can be repeated for the strong coupling rays $(\lambda_x,\lambda_y, \lambda_z)\rightarrow\lambda(\pm1, \mp1, -1)$. Now the $\Phi_\sigma$ field develops the gap, and the sign of $\lambda_p$ determines whether the $x$ or $y$ channel survives. 

\section{Derivation of Renormalization Group Equations}
\label{AppSec:DerivationRGFlowEquations}

In this section, we derive the leading RG equation, Eq.~\eqref{eq:tDeltaFlow}, governing the dipole interactions, tunneling, and CAR processes in the Hamiltonian \eqref{eq:tDeltaBosonized}. We illustrate the general procedure using the $t_0$ tunneling channel,
\begin{equation}
H_{t_0} = \frac{\mathscr t_0 v_\sigma}{4\pi a^2} \int \dd x\cos(\Theta_\sigma)\sin(3\Theta_\rho).
\label{eq:tunnelin0Channel}
\end{equation}
The calculation proceeds in two steps. First, we determine the flow of the dipole couplings and subsequently evaluate their influence on the scaling of the neutral operator $\cos(\Theta_\sigma)$. Finally, we restore the contribution from the charge sector to obtain the flow of the full tunneling operator.

We retain terms of order $\mathcal O(\lambda_i\lambda_j)$ and $\mathcal O(\lambda_i \mathscr t_0)$. Thus, the calculation incorporates the leading corrections beyond tree level while remaining linear in the tunneling amplitude $\mathscr t_0$. Owing to the parametric smallness of $\mathscr t_0$, contributions quadratic in $\mathscr t_0$, as well as their feedback on the charge and neutral interactions, are neglected.

For vanishing $\mathscr t_0$, the expectation value factorizes into charge and neutral contributions. Consequently, the scaling dimension is the sum of the scaling dimensions of the individual factors:
\begin{equation}
\Delta_{t_0} = \Delta_{\sin(3\Theta_\rho)} + \Delta_{\cos(\Theta_\sigma)}.
\label{eqApp:DecomposedScalling}
\end{equation}
In the charge sector, the scaling dimension of $\sin(3\Theta_\rho)$ can be determined using standard methods \cite{GiamarchiBook2003}, yielding $\Delta_{\sin(3\Theta_\rho)} = 3K_\rho/2$. The nontrivial part of the calculation is therefore the interaction correction to the neutral sector.

We work in the imaginary-time formalism and introduce the complex coordinates
\begin{equation}
z = v_\sigma \tau +ix, \quad \bar z=v_\sigma \tau - ix.
\end{equation}
The integration measure is given by
\begin{equation}
d^2 z = \frac{i}{2}\dd z\wedge\dd{\bar z} = v_\sigma \dd\tau\dd x,
\end{equation}
where the wedge product specifies the chosen orientation. According to the $K$ matrix $K_\text{KFP}$ (Eq.~\eqref{eq:KFPmatix}), the neutral mode at the top edge is holomorphic, whereas the neutral mode at the bottom edge is anti-holomorphic. The corresponding free correlation functions are
\begin{subequations}
\begin{align}
\braket{\phi _t(z)\phi_t(w)} & = - \log\left(\frac{z-w}{a}\right), \\
\braket{\phi _b(\bar z)\phi_b(\bar w)} & = - \log\left(\frac{\bar z-\bar w}{a}\right).
\end{align}
\end{subequations}

To isolate the neutral sector, we introduce the simplified Hamiltonian
\begin{multline}
    H_s = 2\pi \sum_{i=x,y,z} \lambda_i \int \dd x \,\mathscr O_i(x) + \frac{\mathscr c}{4\pi a^2}\int \dd x\, \mathscr V(x),
\end{multline}
where 
\begin{equation}
    \mathscr O_i = J_t^iJ_b^i,\qquad \mathscr V = \begin{cases}
        \cos(\Theta_\sigma)\\
        \sin(\Theta_\sigma)\\
        \cos(\Phi_\sigma)\\
        \sin(\Phi_\sigma)
    \end{cases}. \label{eq:VDef}
\end{equation}
The imaginary time, interaction-picture time evolution operator is
\begin{equation}
    \text{S} = \mathcal T_\tau \exp\left(-\int_0^\beta \dd\tau H_s(\tau) \right),
\end{equation}
where $\beta$ is the inverse temperature and $\mathcal T_\tau$ is the imaginary time ordering operator. The series expansion of this operator is then 
\begin{multline}
    \text{S} = 1 - \int d^2 z\, H_s(z,\bar z) \\ + \int_{\tau_z >\tau_w } d^2 z d^2 w \,H_s(z,\bar z) H_s(w,\bar w) + \dots . 
    \label{eqApp:Sexpansion}
\end{multline}
The second-order term is integrated over ordered times (i.e. $\tau_z> \tau_w$). Equivalently, one could
integrate over both time orderings and include a factor 1/2. We use the ordered form throughout. Omitting the term quadratic in $\mathscr t_0^2$, the second-order term becomes
\begin{equation}
\begin{split}
    \text{S}^{(2)}  = & (2\pi)^2\sum_{i,j=x,y,,z}\lambda_i\lambda_j\int_{\tau_z >\tau_w} d^2zd^2 w\, \mathscr O_i(z,\bar z)\mathscr O_j(w,\bar w)\\ & + \sum_i \frac{\lambda_t\mathscr t_0}{2a^2} \int_{\tau_z >\tau_w} d^2zd^2 w\big(\mathscr O_i(z,\bar z)\mathscr V(w,\bar w)\\ &\qquad\qquad\qquad\qquad\qquad\qquad+\mathscr V(z,\bar z)\mathscr O_i(w,\bar w)\big).
\end{split}
\label{eqApp:secondOrderSMatrix}
\end{equation}

In an infinitesimal RG step, we increase the short-distance cutoff from $a$ to $ae^{d\ell}$ by integrating out separations between $a$ and $ae^{d\ell}$, where $d\ell\ll 1$. We decompose the integration domain into a short-distance shell and a long-distance region:
\begin{multline}
\int_{\tau_z >\tau_w} d^2zd^2w \\ = \int_{\substack{\tau_z >\tau_w \\a<|z-w|<ae^{dl}}} d^2zd^2w + \int_{\substack{\tau_z >\tau_w \\ |z-w|>ae^{dl}}} d^2zd^2w.
\end{multline}
Within the short-distance shell, the operators are sufficiently close in the complex plane that their product can be replaced by its operator product expansion. The resulting local terms can then be absorbed into renormalized couplings of the long-distance theory.

Logarithmic corrections to the coupling arise from terms with a $1/|z-w|^2$ prefactor. Introducing the relative coordinate $r=z-w=\rho e^{i\varphi}$, the time-ordering condition $\tau_z>\tau_w$ translates into $\operatorname{Re}(r)>0$. The short-distance shell integral is therefore
\begin{multline}
\int_{\substack{\tau_z>\tau_w \\ a<|z-w|<ae^{d\ell}}} d^2zd^2w\frac{1}{|z-w|^2} \\
= \int_a^{ae^{d\ell}} \frac{d\rho}{\rho}
\int_{-\pi/2}^{\pi/2}d\varphi
= \pi d\ell.
\end{multline}

We first apply the procedure to the dipole coupling, corresponding to the first term in Eq.\eqref{eqApp:secondOrderSMatrix}. For the current operators defined in Eq.\eqref{eq:bosonicFieldRepOfCurrents}, the operator product expansion dictated by the $\mathfrak{su}(2)_1$ algebra reads
\begin{equation}
J^i_t(z)J^j_t(w)
=
\frac{\delta_{ij}}{8\pi^2(z-w)^2}
+
\frac{i\epsilon_{ijk}}{2\pi(z-w)}J_t^k(w),
\end{equation}
with the corresponding expression for the antiholomorphic sector at the bottom edge. Although the bosonic fields on opposite edges obey nontrivial commutation relations in our construction, the current operators at the top and bottom edges still commute. We may therefore regroup the current operators to obtain
\begin{equation}
\begin{split}
\mathscr O_i&(z,\bar z)\mathscr O_j(w,\bar w)
=
J^i_t(z)J_b^j(w)J^i_t(\bar z)J_b^j(\bar w) \\
&=
\frac{\delta_{ij}}{(8\pi^2)^2|r|^4}\mathbf 1
-
\frac{1}{4\pi^2|r|^2}
\sum_{k,l}
\epsilon_{ijk}\epsilon_{ijl}
J^k_t(w)J^l_b(\bar w).
\end{split}
\end{equation}
The constant term proportional to $\delta_{ij}$ shifts the vacuum energy and therefore does not renormalize the couplings considered here. For $i\neq j$, only terms with $k=l$ survive. Combining this result with Eq.~\eqref{eqApp:secondOrderSMatrix} eventually yields
\begin{equation}
\begin{split}
&(2\pi)^2\sum_{i,j}\lambda_i\lambda_j
\int_{\substack{\tau_z>\tau_w\\a<|z-w|<ae^{d\ell}}}
d^2zd^2w
\mathscr O_i(z,\bar z)\mathscr O_j(w,\bar w) \\
&\qquad =
-2\pi d\ell\int d^2w
\left(
\lambda_y\lambda_z\mathscr O_x
+
\lambda_z\lambda_x\mathscr O_y
+
\lambda_x\lambda_y\mathscr O_z
\right).
\end{split}
\end{equation}
To extract the renormalization of the couplings, we compare this result with the first-order term in the S matrix. For instance, for $\lambda_x$, we obtain
\begin{equation}
-2\pi\lambda_x'\int d^2w\,\mathscr O_x
\equiv
-2\pi\left(\lambda_x+\lambda_y\lambda_zd\ell\right)
\int d^2w\,\mathscr O_x.
\end{equation}
Each current operator has a scaling dimension of one, such that the scaling of $\mathscr O_i$ exactly compensates the scaling of the integration measure when rescaling to the original short distance cutoff $a$. Therefore, no additional tree level contributions appear. In the limit $d\ell\rightarrow0$, we obtain for the three couplings the flow equations
\begin{equation}
    \frac{d\lambda_x}{d\ell} = \lambda_y\lambda_z,\quad\frac{d\lambda_y}{d\ell} = \lambda_z\lambda_x,\quad \frac{d\lambda_z}{d\ell} = \lambda_x\lambda_y.
\end{equation}

We now turn to the contributions to coupling $\mathscr c$. Here, it is convenient to use the spinor representations
\begin{subequations}
\begin{align}
    \cos(\Theta_\sigma) &  = \frac{e^{-i\pi/4}}{2} \mathbf z^T_t \sigma_x \mathbf z_b, \\
    \sin(\Theta_\sigma) & = \frac{e^{-i\pi/4}}{2}\mathbf{z}^T_t\sigma_y \mathbf z_b, \\
    \cos(\Phi_\sigma) & = \frac{e^{i\pi/4}}{2}\mathbf z^T_t\sigma_0 \mathbf z_b, \\
    \sin(\Phi_\sigma) & = \frac{e^{-i\pi/4}}{2}\mathbf z_t^T \sigma_z \mathbf z_b.
\end{align}
\end{subequations}
where $\mathbf z_{t/b}$ are the spinor primary fields in Eq.~\eqref{eq:SpinorRep}. 

In the following, we use the leading terms in the operator product expansions between the spinor fields and the current operators. For the top edge, these can be summarized as 
\begin{subequations}
\begin{align}
J^i_t(z) \mathbf z_t(w) \sim \frac{1}{4\pi(z-w)}\mathbf \sigma_i^T\mathbf z_t(w) 
\end{align} 
\end{subequations}
with analogous expressions for the antiholomorphic sector at the bottom edge. We then calculate $\sum_{i=x,y,z} \lambda_i\mathscr O_i(z,\bar z)\mathscr V(w,\bar w) $ for any of the four $\mathscr V$ defined in Eq.~\eqref{eq:VDef} and obtain
\begin{widetext}
\begin{equation}
\begin{split} 
&\sum_{i=x,y,z}\lambda_i\,J^i_t(z)J_b^i(\bar z)\left( \frac{\gamma_\alpha}{2}\mathbf z^T_t(w)\sigma_\alpha \mathbf z_b(w)\right) = \frac{\gamma_\alpha}{2}\sum_{i=x,y,z} \alpha_i\lambda_i\, J^i_t(z)\mathbf z_t^T(w)\sigma_\alpha J^i_b(\bar z) \mathbf z_b(\bar w) \\ & = \frac{\gamma_\alpha}{32\pi^2|z-w|^2}\sum_{i=x,y,z} \alpha_i\lambda_i\,\mathbf z_t^T(w)\sigma_i\sigma_\alpha\sigma_i^T\mathbf z_b(\bar w). \end{split}
\label{eqApp:fusionOandV}
\end{equation} 
\end{widetext} Here, $\gamma_{x,y,z}=e^{- i\pi/4}$ and $\gamma_0=e^{i\pi/4}$ is a phase determined by the precise form of $\mathscr V(w,\bar w)$. The factor $\alpha_i$ is an additional phase arising from moving $J_b(\bar z)$ past the spinor $\mathbf z(w)$, with $\alpha_{x,y}=-1$ and $\alpha_z=1$. Each $\lambda_i$ contributes exactly with the same pre-factor, whereas the sign is determined by the product $\alpha_i \sigma_i \sigma_\alpha\sigma_i^T$. 

For instance, for $\mathscr V(z,\bar z)=\cos(\Theta_\sigma(z,\bar z))$ we find 
\begin{multline}
    \mathscr O_i(z,\bar z)\mathscr V(w,\bar w) +\mathscr V(z,\bar z)\mathscr O_i(w,\bar w) = 2 \mathscr O_i(z,\bar z)\mathscr V(w,\bar w) \\ = -\frac{1}{8\pi^2|z-w|^2}\cos(\Theta(w,\bar w)).
\end{multline}
The short-distance shell integral will contribute another factor of $\pi d\ell$. Using $\alpha=x$ and comparing the such calculated correction with the first order term in Eq.~\eqref{eqApp:Sexpansion} yields
\begin{multline}
    -\frac{\mathscr c'}{4\pi a^2}\int d^2 z\cos(\Theta_\sigma)\\ = - \frac{\mathscr c}{4\pi}\left(1+ \frac{\lambda_x+\lambda_y+\lambda_z}{4}d\ell \right)\int d^2z \cos(\Theta_\sigma).
\end{multline}
When rescaling to the original short distance cutoff $a$, the operator $\cos(\Theta_\sigma)$ together with the integration measure $d^2 z$ contributes an additional contribution of $3 d\ell/2$. The coupled flow equations are therefore
\begin{equation}
    \frac{d\mathscr c}{d\ell} = \left(\frac{3}{2}+\frac{\lambda_x+\lambda_y + \lambda_z}{4}\right)\mathscr c. 
\end{equation}
Using Eq.~\eqref{eqApp:DecomposedScalling}, we retain the the flow equations for the $t_0$ tunneling
\begin{equation}
    \frac{d\mathscr t_0}{d\ell} = \left(\frac{3}{2}-\frac{3}{2}K_\rho + \frac{\lambda_x +\lambda_y + \lambda_z}{4}\right)\mathscr t_0.
\end{equation}
This is the origin of Eq.~\eqref{eq:RGt0} in the main text. 

Repeating the calculation with choosing $\alpha=x,y,z$ in Eq.~\eqref{eqApp:fusionOandV} reproduces remaining flow equations \eqref{eq:RGFlowTx}-\eqref{eq:RGtz}.

\section{Disorder Gauge Transformation}
\label{AppSec:disorderGaugeTrafo}

The $\mathfrak{su}(2)_1$ current algebra is invariant under local orthogonal $\mathrm{SO}(3)$ transformations,
\begin{equation}
J^i_\tau = S^{ij}_\tau \tilde J^{j}_\tau + s_\tau h_{\tau}^i,
\end{equation}
where
\begin{equation}
S_\tau = \mathcal T_x \exp\left(\int_{x_0}^x \dd{x'} M_\tau(x') \right),
\end{equation}
and
$[S_\tau \partial_x S_\tau^T]_{ab} = 4\pi \epsilon_{abi} h^i_\tau.$
For the matrix $M_\tau(x)$, we use the standard basis in which the generators of $\mathrm{SO}(3)$ are given by $(L_i)_{ab} = -\epsilon_{abi}$. We then find
\begin{equation}
M_\tau(x) = 4\pi \sum_{i=x,y,z} h^i_\tau(x) L_i.
\end{equation}
To determine the coefficients $h^i_\tau$ that absorb the disorder into the current basis, we transform the Hamiltonian $H_\sigma + H_\xi$ and choose $h^i_\tau$ such that all terms linear in $J^i_\tau$ vanish. This procedure yields the system of equations
\begin{subequations}
\begin{align}
\frac{2}{3}h_t^i - \lambda_i h_b^i + \frac{\xi^x_t}{v_\sigma}\delta_{xi} + \frac{\xi^y_t}{v_\sigma}\delta_{yi} &= 0, \\
-\frac{2}{3}h_b^i + \lambda_i h_t^i + \frac{\xi^x_b}{v_\sigma}\delta_{xi} + \frac{\xi^y_b}{v_\sigma}\delta_{yi} &= 0.
\end{align}
\end{subequations}
Solving this system gives
\begin{equation}
h_\tau^i =
\begin{cases}
s_\tau\dfrac{\frac{2}{3}\xi_\tau^i-\lambda_i \xi^i_{\bar\tau}}{v_\sigma\left(\lambda_i^2 - (2/3)^2\right)}, & \text{for } i=x,y, \\
0, & \text{for } i=z,
\end{cases}
\end{equation}
where $\bar\tau$ denotes the edge opposite to $\tau$. For the special case of edge-correlated disorder, $\xi_t = \xi_b^* \equiv \xi$, we have $\xi_t^x = \xi_b^x$ and $\xi_t^y = -\xi_b^y \equiv \xi^y$. The coefficients $h_\tau^i$ then simplify considerably, yielding the following $M$-matrices:
\begin{subequations}
\begin{align}
M_t &= -\frac{4\pi\xi^x}{v_\sigma\left(\lambda_x + \frac{2}{3}\right)} L_x
+\frac{4\pi\xi^y}{v_\sigma\left(\lambda_y - \frac{2}{3}\right)} L_y, \\
M_b &= \frac{4\pi\xi^x}{v_\sigma\left(\lambda_x + \frac{2}{3}\right)} L_x+
\frac{4\pi\xi^y}{v_\sigma\left(\lambda_y - \frac{2}{3}\right)} L_y.
\end{align}
\end{subequations}
Thus, $M_t$ and $M_b$ are nearly identical, differing only by a relative sign in the $x$ component. This observation establishes the relation
\begin{equation}
S_t = \eta S_b \eta, \qquad \text{where} \qquad
\eta = \begin{pmatrix}
-1 & 0 & 0 \\
0 & 1 & 0 \\
0 & 0 & -1
\end{pmatrix}.
\end{equation}

In contrast to the currents, the fermion bilinears, which are proportional to the matrix $g$, transform under an SU(2) matrix constructed from $S_\tau$:
\begin{equation}
U_\tau = \mathcal{T}_x \exp\left(2\pi i \int_{x_0}^x \dd{x'} \left[(-1)^\tau h^x \sigma_x - h^y \sigma_y\right]\right).
\end{equation}
From this, we can deduce the important relation
$ U_b = \sigma_y U_t \sigma_y $
for correlated disorder. Additionally, because of the Klein factor in the fermions, $U_\tau$ must be rotated accordingly, yielding
\begin{multline}
\hat{U}_\tau = D_\kappa U_\tau D_\kappa
\\
= \mathcal{T}_x \exp\left(2\pi i \int_{x_0}^x \dd{x'} \left[(-1)^\tau h^x \sigma_y + h^y \sigma_x\right]\right),
\label{eqApp:DefinitionHatU}
\end{multline}
and
$\hat{U}_t = \sigma_x \hat{U}_b \sigma_x.$

The matrices $S_\tau(x)$ and $\hat U_\tau(x)$ are random matrices. They can be understood as being generated by a random walk through the corresponding Lie group, driven by the zero-mean Gaussian random variables
\begin{subequations}
\begin{align}
\overline{h^x(x)h^x(x')} &= D_x\delta(x-x'), \quad
D_x = \frac{W}{2v_\sigma^2(\lambda_x+2/3)^2}, \\
\overline{h^y(x)h^y(x')} &= D_y\delta(x-x'), \quad
D_y = \frac{W}{2v_\sigma^2(\lambda_y-2/3)^2}.
\end{align}
\label{eqApp:DxDyCorrelations}
\end{subequations}
The random walk is parameterized by the $x$ coordinate, starts at $x_0$, where, by definition, $S_\tau(x_0)=U_\tau(x_0)=\mathbf{1}$, and terminates at $x$. For a sufficiently long path $[x_0,x]$, the matrices $S_\tau(x)$ and $U_\tau(x)$ lose their memory of the initial state and can be treated as uniformly distributed over the corresponding Lie group.
More formally, as the path length $x-x_0$ tends to infinity, the probability distributions $P(U_\tau)$ of $U_\tau(x)$ converge weakly to the normalized Haar measure on the compact Lie group, independently of the initial state \cite{ChirikjianBook2012}. In particular,
\begin{equation}
P(U_\tau)\mathcal{D}U_\tau \xrightarrow{x_0\to-\infty} \mathcal{D}U_\tau,
\end{equation}
where $\mathcal{D}U_\tau$ denotes the normalized Haar measure.
For path lengths greater than $l_\mathrm{dis}\sim v_\sigma^2/W$, the angular range of the accumulated random rotation $U_\tau(x)$ becomes of order unity, and the information about the initial state is lost. We therefore assume that the probability distribution is approximately uniform with respect to the Haar measure and may formally take the initial point $x_0$ to $-\infty$.
The analogous statement holds for $S_\tau(x)$.

\section{Effective Replica Action}

While the transformations $S_\tau$ manage to absorb the disorder Hamiltonian $H_\xi$ into the definitions of the current operators, it will reappear wherever the two edges are coupled. In particular, this is true for the tunneling and pairing.

The tunneling and pairing channels are proportional to $\text{tr}[\sigma_i g]$, where $g$ transforms non-trivially under $\hat U_\tau$. In order to account for how, we derive an effective replica action in this section. 

\subsection{Average of first moments}

In the following calculations, we need to evaluate expectation values of the form $\overline{\hat{U}_t^\dagger{\sigma_i}\hat U^\dagger_b}$ and establish that
\begin{equation}
    \overline{\hat{U}^\dagger_t\sigma_i \hat{U}_b} = \overline{\hat{U}^\dagger_t\sigma_i \sigma_x \hat{U}_t}\sigma_x.
\end{equation} This immediately demonstrates that the $\sigma_x$ channel is unaffected by the edge-correlated disorder.

Using equations \eqref{eqApp:DefinitionHatU} and \eqref{eqApp:DxDyCorrelations}, we find that the expectation values of the remaining channels are 
\begin{subequations} 
\begin{align}    
    \overline{\sigma_0(x)}\equiv\overline{\hat U_t^\dagger \sigma_0 \hat U_b} & = e^{-8\pi^2D_x(x-x_0)}\sigma_0, \\ \overline{\sigma_y(x)}\equiv\overline{\hat U_t^\dagger \sigma_y \hat U_b} & = e^{-8\pi^2(D_x+D_y)(x-x_0)}\sigma_y, \\ \overline{\sigma_z(x)}\equiv\overline{\hat U_t^\dagger \sigma_z \hat U_b} & = e^{-8\pi^2D_y(x-x_0)}\sigma_z.
\end{align} 
\label{eqApp:firstMomentAverage}
\end{subequations} In the limit $x_0\rightarrow-\infty$, all channels except the $x$ channel average to zero.

\subsection{Average of the Second Moment}

We will also need the average over the second moment of the SU(2) matrix-valued random variables $\sigma_\alpha(x)$ of the shape
$\overline{\text{tr}[A\sigma_\alpha(x)]\text{tr}[B\sigma_\beta(x')]}$,
where $A$ and $B$ are arbitrary $2\times 2$ matrices.
Since the disorder on disjoint intervals is statistically independent, the average over $x_0\leq x \leq x'$ can be evaluated in two steps. First, we average over disorder between $x$ and $x'$ and later take the average over the remaining interval between $x_0$ and $x$. To make this explicit, we define the interval average
\begin{align}
    \braket{F}_{[a,b]} & = \frac{1}{Z_{[a,b]}}\int \mathcal Dh_{[a,b]}P_{[a,b]}(h) F, \\ Z_{[a,b]} & =\int \mathcal Dh_{[a,b]}P_{[a,b]}(h),
\end{align}
where $h = \{h^i\}_{i=x,y}$ is the collection of the random fields defined in Eq.~\eqref{eqApp:DxDyCorrelations} and the subscript $[a,b]$ denotes a restriction onto the latter interval. Further, we separate $\sigma_\alpha(x') = \hat U^\dagger_t(x',x)\sigma_i(x)\hat U_b(x',x)$, where 
\begin{equation}
    \hat{U}_\tau(b,a) = \mathcal T_x \exp\left(2\pi i \int_{a}^{b} \dd {x''}(-1)^\tau h^x\sigma_y+h^y\sigma_x\right).
\end{equation}
is the generalization of Eq.~\eqref{eqApp:DefinitionHatU} for the arbitrary interval $[a,b]$. Since the rotation $\hat U_\tau(x,x')$ depends only on disorder in the interval [x,x'] may therefore write
\begin{multline}
    \overline{\text{tr}[A\sigma_\alpha(x)]\text{tr}[B\sigma_\beta(x')]} \\
    = \left\langle\text{tr}[A\sigma_\alpha(x)]\text{tr}[B\big\langle U^\dagger_t(x',x)\sigma_\beta(x) U(x',x)\big\rangle_{[x_,x']}]\right\rangle_{[x_0,x]}
\end{multline}
The inner average over the interval $[x,x']$ can be read off from Eq.~\eqref{eqApp:firstMomentAverage} and we find
\begin{equation}
    \big\langle U^\dagger_t(x',x)\sigma_\beta(x) U(x',x)\big\rangle_{[x_,x']} = e^{-\frac{W}{\gamma_\beta v_\sigma^2}(x'-x)}\sigma_\beta(x), 
\end{equation}
where 
\begin{subequations}
    \begin{align}
        \gamma_0 & = \left(\frac{\lambda_x+2/3}{2\pi}\right)^2, \\
        \gamma_x & = \infty, \\
        \gamma_y & = \left[\left(\frac{2\pi}{\lambda_x+2/3}\right)^2 + \left(\frac{2\pi}{\lambda_y-2/3}\right)^2\right]^{-1}, \\
        \gamma_z & =\left(\frac{\lambda_y-2/3}{2\pi}\right)^2.
    \end{align}
\end{subequations}
In particular, this translates to 
\begin{multline}
    \overline{\text{tr}[A\sigma_\alpha(x)]\text{tr}[B\sigma_\beta(x')]} \\= e^{-\frac{W}{\gamma_\beta v_\sigma^2}(x'-x)}\overline{\text{tr}[A\sigma_\alpha(x)]\text{tr}[B\sigma_\beta(x)]}.
\end{multline}
Now the average is only over a local expression on the right-hand side of the equal sign. For $x-x_0> l_\text{dis}\sim v_\sigma^2/W$ the matrices $U_\tau(x)$ are uniformly distributed over SU(2) and Weingarten calculus applies:
\begin{widetext}
\begin{equation}
    \overline{\text{tr}[A\sigma_\alpha(x)]\text{tr}[B\sigma_\beta(x)]} = \begin{cases}
        \frac{2}{3}\left(\text{tr}[\sigma_xA\sigma_xB]-\frac{1}{2}\text{tr}[\sigma_x A]\text{tr}[\sigma_x B]\right), & \alpha=\beta =0 \\ \text{tr}[\sigma_x A]\text{tr}[\sigma_x B], & \alpha=\beta =x \\
        -\frac{2}{3}\left(\text{tr}[\sigma_xA\sigma_xB]-\frac{1}{2}\text{tr}[\sigma_x A]\text{tr}[\sigma_x B]\right), & \alpha=\beta \in \{y,z\} \\
        0 & \alpha\neq\beta  
    \end{cases}
    \label{eqApp:secondMomentWeingartenAverage}
\end{equation}
\end{widetext}

The same arguments also hold for $x>x'$, and since the average over the local part is diagonal in $\alpha$ and $\beta$ we can summarize both results in the statement
\begin{multline}
    \overline{\text{tr}[A\sigma_\alpha(x)]\text{tr}[B\sigma_\beta(x')]} \\ = e^{-\frac{W}{\gamma_\alpha v_\sigma^2}|x'-x|}\overline{\text{tr}[A\sigma_\alpha(x)]\text{tr}[B\sigma_\beta(x)]}.
    \label{eqApp:secondMomentNonLocal+Local}
\end{multline}

If the typical fluctuations of the bosonic fields happen on length scales larger than $l_\text{dis}$, we can approximate the average to be approximately uncorrelated at different positions and write 
\begin{equation}
    e^{-\frac{W}{\gamma_iv_\sigma^2}|x-x'|}\approx \gamma_i \frac{2v_\sigma^2}{W}\delta (x-x')
    \label{AppEq:DeltaApproximation}
\end{equation}
which holds in a distributional sense.

\subsection{Derivation of Replica Action}
\label{secApp:replicaAction}

As a first step, we lift the Hamiltonian \eqref{eq:fullHamiltonian} to the action
\begin{equation}
\begin{split}
    \mathcal S & = \frac{3}{4\pi} \int \ddd x\tau \begin{pmatrix}
        \partial_\tau\Phi_\rho & \partial_\tau\Theta_\rho
    \end{pmatrix} 
    \begin{pmatrix}
        0 & 1 \\ 1 & 0
    \end{pmatrix}
    \begin{pmatrix}
        \partial_x \Phi_\rho \\ \partial_x \Theta_\rho 
    \end{pmatrix}
    \\ & + \frac{1}{4\pi} \int \ddd x\tau \begin{pmatrix}
        \partial_\sigma\Phi_\sigma & \partial_\tau\Theta_\sigma
    \end{pmatrix} 
    \begin{pmatrix}
        0 & -1 \\ -1 & 0
    \end{pmatrix}
    \begin{pmatrix}
        \partial_x \Phi_\sigma \\ \partial_x \Theta_\sigma 
    \end{pmatrix} \\
    & + \int \dd \tau H(\tau).
\end{split}
\end{equation}
After applying the rotations $S_\tau(x)$ and $\hat{U}_\tau(x)$, the contribution $H_\xi$ is absorbed into the field definitions of $\tilde \Theta_{p}$ and $\tilde \Phi_p$, couplings in $H_\sigma$, $H_t$ and $H_\Delta$ which couple the top and bottom edge are disorder dependent now.

To deal with the disorder, we take the disorder average
\begin{multline}
    \overline{\tilde H_\sigma^{tb}} = 2\pi v_\sigma \int \dd x \widetilde{\v J}^{\text{T}}_t \overline{S_t^\text{T}(x)\,\Lambda\, S_b(x)}\widetilde {\v J}_b \\ = 2\pi v_\sigma \int \dd x \overline{\text{tr}\left[S_t^\text{T}(x) \Lambda S_b(x)(\widetilde{\v J}_b \otimes \widetilde{\v J}_t^\text{T})\right]}
\end{multline}
of the dipole interaction, 
where $\Lambda = \text{diag}(\lambda_x, \lambda_y,\lambda_z)$ is a diagonal matrix containing all dipol couplings and $(\widetilde{\v J}_b \otimes \widetilde{\v J}_t^\text{T})$ is a tensor product of the current operators. Using Weingarten calculus for the average yields
\begin{multline}
    \overline{\tilde H_\sigma^{tb}} = \frac{2\pi v_\sigma}{3}\int \dd x \text{tr}[\eta \Lambda]\text{tr}[\eta(\widetilde{\v J}_b \otimes \widetilde{\v J}_t^\text{T})] \\ = 2\pi v_\sigma\bar\lambda\int\dd x \widetilde{\v J}_t^\text{T}\eta\widetilde{\v J}_b.
\end{multline}
Hence, the disorder average projects on the strong coupling ray $\bar\lambda (-1, 1, -1)$, where $\bar \lambda = (-\lambda_x+\lambda_y - \lambda_z)/3$. Even though higher order terms in $\bar \lambda$ should also be considered, we only keep $\overline{\tilde H^{tb}_\sigma}$ since all higher order terms are RG irrelevant in the perturbative regime where $\lambda_{x,y,z}\ll 1$. For instance, the variance $\overline{(\tilde H_\sigma^{tb})^2}-\overline{\tilde H^{tb}_\sigma}^2 \propto \mathscr D_\sigma$ has (on tree level) the scaling dimension $\Delta_\sigma =4$ since it is a product of four current operators. Thus, the disorder coupling obeys the flow equation
\begin{equation}
    \frac{d\mathscr D_\sigma}{d\ell} = (3-4)\mathscr D_\sigma = - \mathscr D_\sigma.
\end{equation}
This is the same argument as the irrelevance of the $\tilde  g^i_{\tau\tau'}$ couplings in Hamiltonian \eqref{eq:FullInterTrenchInteraction} at the KFP fixed point.

We continue with $\tilde H_t$ and $\tilde H_\Delta$. In the following, it is particularly handy to express the bilinears in terms of the field-valued SU(2) matrix $g$. Note that the $x$-channel is disorder independent. It is easily checked when noting that $\hat U_b = \sigma_x \hat U _t \sigma_x$ and, thus, $\hat U^\dagger_t\sigma_x U_b = U_t^\dagger \sigma_x\sigma_x U_b \sigma_x = \sigma_x$. Hence, we separate the $\tilde H_t$ and $\tilde H_b$ in disorder dependent and independent parts. For notational convenience, let the Hamiltonians $\delta\tilde H_t$ and $\delta\tilde H_\Delta$ only contain the disordered $0, y, z$ channels, that is 
\begin{multline}
    \delta \tilde H_t =  \frac{\sqrt2 t_0}{\pi a}\int \dd x\text{tr}[\tilde g]\sin(3\tilde\Theta_\rho) \\ + i\frac{\sqrt{2}}{\pi a}\int \dd x \left(t_x\text{tr}[\sigma_y\tilde g] + t_z\text{tr}[\sigma_z\tilde g]\right)\cos(3\Theta_\rho)
\end{multline}
and
\begin{equation}
    \begin{split}
       \delta \tilde H_\Delta & = -\frac{\sqrt{2}\Delta _0}{\pi a}\int \dd x \text{tr}[\tilde g]\sin(3\tilde\Phi_\rho) \\
       & + \frac{\sqrt{2}|d_y|}{\pi a}\int\dd x\text{tr}[\sigma_y\tilde g]\cos(3\tilde\Phi_\rho + \gamma_y) \\
       & + \frac{\sqrt{2}|d_z|}{\pi a}\int\dd x\text{tr}[\sigma_z\tilde g]\cos(3\tilde\Phi_\rho + \gamma_z).
    \end{split}
\end{equation}

After introducing the replica fields $\tilde\Theta_p^n$ and $\tilde \Phi_p^n$ taking the average, we find the effective replica action 
\begin{multline}
    \mathcal S_\text{rep} = \sum_{n}\mathcal S_0[\tilde \Theta^n,\tilde \Phi^n] \\ - \sum_{nm}\int \text{d}\tau\text{d}\tau' \overline{(\delta\tilde H^n_t+\delta\tilde H^n_\Delta)(\delta\tilde H'^m_t+\delta\tilde H'^m_\Delta)}.
\end{multline}
where $\delta H^n_{t,\Delta}=\delta H^n_{t,\Delta}(\tau)$ and $\delta H'^m_{t,\Delta}=\delta H'^m_{t,\Delta}(\tau')$. Technically, one would need to take all moments into account as the probability is non-Gaussian. However, we stop at the first non-trivial order in $t_\alpha$ and $d_0, d_i$ since they are assumed to be parametrically small. Also, any higher order in the average would be more RG irrelevant then the second order. All disorder-independent contributions are collected in $\mathcal S_0$. Note that $\overline{\delta \tilde H^n_{t,\Delta}}=0$ since $\overline{\hat U^\dagger_t g U_b}= 0$ according to Weingarten calculus.

The second-order terms can be calculated using Eqs.~\eqref{eqApp:secondMomentWeingartenAverage}, ~\eqref{eqApp:secondMomentNonLocal+Local}, and ~\eqref{AppEq:DeltaApproximation}. Starting with the tunneling part, we get
\begin{equation}
\begin{split}
    \, & \overline{\delta \tilde H^n_t \delta \tilde H'^m_t} = \frac{2}{\pi^2 a^2}\int \ddd x {x'} \\ &\Bigg(t_0^2 \overline{\text{tr}[\hat U_b\sigma_0\hat U^\dagger_t \tilde g^n]\text{tr}[\hat U_b'\sigma_0\hat U'^\dagger_t \tilde g'^m]}\sin(3\Theta^n_\rho)\sin(3\Theta'^m_\rho) \\
    & + t_y^2 \overline{\text{tr}[\hat U_b\sigma_y\hat U^\dagger_t \tilde g^n]\text{tr}[\hat U_b'\sigma_y\hat U'^\dagger_t \tilde g'^m]} \cos(3\Theta^n_\rho)\cos(3\Theta'^m_\rho)\\
    & + t_z^2 \overline{\text{tr}[\hat U_b\sigma_z\hat U^\dagger_t \tilde g^n]\text{tr}[\hat U_b'\sigma_z\hat U'^\dagger_t \tilde g'^m]}\cos(3\Theta^n_\rho)\cos(3\Theta'^m_\rho)\Bigg). 
\end{split}
\end{equation}
Here, $'$ indicates that the field depends on $x'$ and $\tau'$.
Note that we only kept the diagonal terms as cross terms vanish according to Eq.~\eqref{eqApp:secondMomentWeingartenAverage}.
Let us start with the first $0$ channel, for which we need to calculate the average 
\begin{equation}
\begin{split}
    \overline{A^{nm}_0} & = \overline{\text{tr}[\sigma_{0}\hat U^\dagger_t\tilde g^n \hat U_b]\text{tr}[\sigma_{0}\hat U^\dagger_t\tilde g'^m \hat U_b]} \\ & = \overline{\text{tr}[\sigma_x \tilde g^n \sigma_x(x)]\text{tr}[\sigma_x \tilde g'^m \sigma_x(x)]} \\
    & =\frac{2}{3}\left(\text{tr}[\sigma_x\tilde g^n \sigma_x\tilde g'^m] - \frac{1}{2}\text{tr}[\sigma_x \tilde g^n]\text{tr}[\sigma_x \tilde g'^m]\right),
\end{split}
\end{equation}
where we defined $\sigma_x(x) = \hat U^\dagger_t\sigma_x\hat U_t$ and used Eq.~\eqref{eqApp:secondMomentWeingartenAverage} to average over the random SU(2) matrix $\hat U_t$. The average $\overline{A^{nm}_0}$ expressed in the bosonic fields $\tilde \Phi_{\sigma}$ and $\tilde \Theta_{\sigma}$ yields 
\begin{equation}
    \overline{A^{nm}_0} = \frac{1}{3}\left(\cos(\tilde \Theta_\sigma^n  - \tilde \Theta'^m_\sigma)+ \cos(\tilde \Phi_\sigma^n)\cos(\tilde\Phi'^m_\sigma)\right).
\end{equation}
The calculation can be repeated for the $y$ and $z$ channel and yields $\overline{A^{nm}_0} = -\overline{A^{nm}_{yz}}\equiv \overline{A^{nm}}$ with which we write the tunneling as
\begin{equation}
\begin{split}
    \overline{\delta \tilde H^n_t \delta \tilde H'^m_t}  = \frac{4v_\sigma ^2}{\pi^2 a^2 W} \int \dd x \overline{A^{nm}} \Big( \gamma_0t_0^2 \sin(3 \Theta^n_\rho)\sin(3 \Theta'^m_\rho) \\
    +(\gamma_y t^2_y + \gamma_zt_z^2)\cos(3\Theta^n_\rho)\cos(3\Theta'^m_{\rho})\Big)
\end{split}
\end{equation}
Note that, although we have integrated over $x'$ using the $\delta$-function in Eq.~\eqref{AppEq:DeltaApproximation}, the fields still depend on $\tau'$, and we must therefore retain the primes. At this point, it is customary to use trigonometric identities to rewrite the average in terms of $\cos(3\Theta_\rho^n - 3\Theta'^m_\rho)$ and $\cos(3\Theta_\rho^n + 3\Theta'^m_\rho)$, retaining only the former. The reason is that the replica-diagonal terms of the former renormalize the Luttinger parameter $K_\rho$, whereas those of the latter generate a higher harmonic and are therefore less relevant in the RG sense.

Eventually, we arrive at the final form of the tunneling contribution after disorder averaging:
\begin{multline}
    \overline{\delta \tilde H^n_t \delta \tilde H'^m_t} = \frac{\mathscr D_\Theta}{2\pi a^2}\int \dd x \cos(\tilde \Theta^n_\sigma - \tilde \Theta'^m_\sigma)\cos(3\Theta_\rho^n - 3\Theta'^m_\rho) \\ +
    \frac{\mathscr D_\Phi}{2\pi a^2}\int \dd x \cos(\tilde\Phi^n_\sigma)\cos(\tilde\Phi'^m_\sigma)\cos(3\Theta_\rho^n - 3\Theta'^m_\rho),
\end{multline}
where 
\begin{equation}
    \mathscr D_{\Theta} = \mathscr D_\Phi = \frac{4v_\sigma^2}{\pi W}\Gamma\bar t^2,
\end{equation}
with $\bar t = \sqrt\frac{\gamma_0t_0^2 + \gamma_xt_x^2 + \gamma_zt_z^2}{\gamma_0+\gamma_y+\gamma_z}$ being the root square mean of all disorderd coupling constants, and $\Gamma = \gamma_0 + \gamma_y + \gamma_z$ is the total decay rate of spatial disorder correlations and depends on $\lambda_x$ and $\lambda_y$. Even though the effective disorder strengths $\mathscr D_\Phi$ and $\mathscr D_\Theta$ have the same bare value, under renormalization, they are renormalized differently and must thus be treated separately.

The former derivation can also be one-to-one applied to the average over the pairing and the mixed terms. For the CAR. We get 
\begin{multline}
    \overline{ \delta\tilde H_\Delta^n\delta \tilde H'^m_\Delta} = \frac{\mathscr C_\Theta}{2\pi a^2}\int \dd x \cos(\tilde \Theta_\sigma^n - \tilde\Theta'^m_\sigma)\cos(3\Phi_{\rho}^n - 3\Phi'^m_{\rho})  \\ 
    \frac{\mathscr C_\Phi}{2\pi a^2}\int \dd x\cos(\tilde\Phi_\sigma^n)\cos(\tilde \Phi'^m_\sigma)\cos(3\Phi_{\rho}^n - 3\Phi'^m_{\rho}),
\end{multline}
where 
\begin{equation}
    \mathscr C_\Theta = \mathscr C_\Phi = \frac{4v_\sigma^2}{\pi W}\Gamma \bar \Delta^2
\end{equation}
with the equivalent root square mean of the pairing $\bar \Delta = \sqrt\frac{\gamma_0 d_0^2 + \gamma_y|d_y|^2+\gamma_z|d_z|^2}{\gamma_0 + \gamma_y + \gamma_z}$. 

The mixed terms of the shape $\overline{\delta \tilde H_t^a \delta \tilde H'^b_\Delta}$ can be derived similarly with the couplings $\mathscr F_\Phi$ and $\mathscr F_\Theta$. However, in the following, we neglect those as they are less relevant in the low-energy theory. This can be simply understood by comparing the scaling dimensions (in the weak exchange case) of the respective operators:  
\begin{subequations}
\begin{align}
    \Delta_{\mathscr D_\Theta} & = 3K_\rho -1+\frac{\lambda_z}{2}, \\
    \Delta_{\mathscr D_\Phi} & = 3K_\rho -1-\frac{\lambda_z}{2}, \\
    \Delta_{\mathscr C_\Theta} & = \frac 3 {K_\rho} -1+\frac{\lambda_z}{2}, \\
    \Delta_{\mathscr C_\Phi} & = \frac 3 {K_\rho}  -1-\frac{\lambda_z}{2}, \\
    \Delta_{\mathscr F_\Theta} & = \frac{3}{2}\left(K_\rho + \frac{1}{K_\rho}\right) -1+\frac{\lambda_z}{2}, \\
    \Delta_{\mathscr F_\Phi} & = \frac{3}{2}\left(K_\rho + \frac{1}{K_\rho}\right) -1-\frac{\lambda_z}{2}.
\end{align}
\end{subequations}
Thus, the mixed terms have for almost all $K_\rho$, but for the special case $K_\rho=1$, there is at least one operator with a larger scaling dimension and are, that is 
\begin{equation}
\text{min}(\Delta_{\mathscr C_{\Phi,\Theta}}, \Delta_{\mathscr D_{\Phi,\Theta}}) \leq \Delta_{\mathscr F_{\Phi,\Theta}} \leq \text{max}(\Delta_{\mathscr C_{\Phi,\Theta}}, \Delta_{\mathscr D_{\Phi,\Theta}})
\end{equation}

 This section is concluded by stating the full low-energy replicated action 
 \begin{widetext}
 \begin{equation}
 \begin{split}
     \mathcal S_\text{rep} & = \sum_n \mathcal S_0[\tilde \Theta^n, \tilde \Phi^n] -\frac{2t_x}{\pi a} \sum_n \int \ddd{x}{\tau} \sin(\tilde\Phi^n_\sigma)\cos(3\tilde\Theta^n_\rho) - \frac{2|d_x|}{\pi a}\sum_n\int \ddd{x}{\tau}\sin(\tilde \Phi^n_\sigma)\cos(3\tilde\Phi^n_\rho + \delta_x) \\ &  - \frac{\mathscr D_\Theta}{2\pi a^2}\sum_{nm}\int\text{d}\tau\text{d}\tau'\text{d}x \cos(\tilde \Theta^n_\sigma - \tilde\Theta'^m_\sigma)\cos(3 \Theta^n_\rho - 3\Theta'^m_\rho) 
      - \frac{\mathscr D_\Phi}{2\pi a^2}\sum_{nm}\int\text{d}\tau\text{d}\tau'\text{d}x \cos(\tilde \Phi^n_\sigma)\cos(\tilde \Phi'^m_\sigma)\cos(3 \Theta^n_\rho - 3\Theta'^m_\rho) \\ &
     - \frac{\mathscr C_\Theta}{2\pi a^2}\sum_{nm}\int\text{d}\tau\text{d}\tau'\text{d}x \cos(\tilde \Theta^n_\sigma - \tilde\Theta'^m_\sigma)\cos(3 \Phi^n_\rho - 3\Phi'^m_\rho)  - \frac{\mathscr C_\Phi}{2\pi a^2}\sum_{nm}\int\text{d}\tau\text{d}\tau'\text{d}x \cos(\tilde \Phi^n_\sigma)\cos(\tilde \Phi'^m_\sigma)\cos(3 \Phi^n_\rho - 3\Phi'^m_\rho).
\end{split}
 \end{equation}
 \end{widetext}

\bibliographystyle{apsrev4-2-ten}
\bibliography{references}

\end{document}